\documentclass[aps,prb,groupaddress,twocolumn,floatfix]{revtex4-2}
\usepackage{units}
\usepackage{amsmath}
\usepackage{amssymb}
\usepackage{graphicx}
\usepackage{bm}
\usepackage{multirow,color,relsize,ulem,microtype}
\usepackage{array}
\usepackage{booktabs,cancel}

\newcommand{\be}{\begin{equation}}
\newcommand{\ee}{\end{equation}}

\newcommand{\re}[1]{\text{Re}\left[#1\right]}
\newcommand{\im}[1]{\text{Im}\left[#1\right]}

\DeclareMathOperator{\Tr}{Tr}

\begin{document}

\title{Non-Hermitian gauged laser arrays with localized excitations: Anomalous threshold and generalized principle of selective pumping
}

\author{Li Ge}
\email{li.ge@csi.cuny.edu}
\affiliation{\textls[-18]{Department of Physics and Astronomy, College of Staten Island, CUNY, Staten Island, NY 10314, USA}}
\affiliation{The Graduate Center, CUNY, New York, NY 10016, USA}

\author{Zihe Gao}
\affiliation{Department of Materials Science and Engineering, University of Pennsylvania, Philadelphia, PA 19104, USA}

\author{Liang Feng}
\affiliation{Department of Materials Science and Engineering, University of Pennsylvania, Philadelphia, PA 19104, USA}

\date{\today}

\begin{abstract}
We investigate non-Hermitian skin modes in laser arrays with spatially localized excitation. Intriguingly, we observe an unusual threshold behavior when selectively pumping either the head or the tail of these modes: both cases exhibit the same lasing threshold and hence defy the conventional principle of selective pumping, which aims to maximize the overlap between the pump profile and the target lasing mode. To shed light on this enigma, we reveal a previously overlooked phenomenon, i.e., energy exchange at non-Hermitian coupling junctions with the photonic environment, which does not occur with uniform gain or loss. 
Utilizing a transfer matrix approach, we elucidate the mechanism of this anomalous threshold behavior, which is determined by the specific physical realization of the non-Hermitian gauge field (i.e., using gain, loss, or their mixture). 
Finally, we derive a generalized principle of selective pumping in non-Hermitian arrays, which shows that the decisive spatial overlap is given by the triple product of the pump, the lasing mode, and its biorthogonal partner. 
Our study provides a glimpse into how the two forms of non-Hermiticity, i.e., asymmetric couplings and a complex onsite potential, interact synergetically in laser arrays, which may stimulate further explorations of their collective effects in photonics and related fields. 

\end{abstract}

\maketitle

\section{Introduction}

Spatially selective pumping is a widely employed technique in driven-dissipative systems, where the excitation energy, or ``pump,'' is concentrated on specific regions of the system under study \cite{Hermite,Hermite2,Woerdman,Narimanov,Wang,rex,Aung,Ge_selective1,spiral,Choi,Seng,Fukushima,Ge_selective2,Seng2,Ge_selective3,Rotter,Sebbah,Sebbah2,Manni,Ge_selective4,Sun}, typically using materials with a broad gain spectrum and an incoherent source at a frequency higher than that of the target mode. By spatially overlapping the pump with the targeted mode, this approach provides a means of single- and few-mode excitation, complementary to spectrally coherent drive, especially in scenarios where the density of states exceeds the inverse spectral resolution of the coherent excitation source. 

Using spatially selective pumping, early experiments in side-pumped solid state lasers demonstrated Hermite-Gaussian modes \cite{Hermite,Hermite2} and ray modes \cite{Woerdman}. Subsequently, the advent of microcavity lasers \cite{Microcavity1,Microcavity2,Microcavity3} has elevated spatially selective pumping to an indispensable tool for exciting and observing wave-chaotic modes \cite{Narimanov,Wang}, reducing lasing threshold \cite{rex,Aung}, enhancing output power \cite{Ge_selective1}, tailoring emission directionality \cite{spiral,Choi,Seng,Fukushima}, and controlling modal interaction and multimode lasing behaviors \cite{Ge_selective2,Seng2,Ge_selective3}.  In addition, spatially selective pumping has also been successfully employed to manipulate properties of random lasers \cite{Rotter,Sebbah,Sebbah2} and induce pattern formations in exciton-polariton condensates \cite{Manni,Ge_selective4,Sun}.

More recently, spatially selective pumping has attracted considerable interest in photonic molecules and lattices. For example, it is fundamental to non-Hermitian photonics based on quantum inspired symmetries \cite{RMP,NPreview}, such as parity-time symmetry that requires judiciously placed elements of optical gain and loss. Non-Hermitian degeneracies known as exceptional points, as well as the phase transitions across them, have been observed using spatially selective pumping \cite{NPhyreview,NMatreview}, showcasing their remarkable sensitivity to changes in system parameters. Coupled with carrier dynamics, spatially selective pumping has also provided a means to tune nonlinear properties of coupled semiconductor (class B) lasers \cite{Kominis,Kominis2,Dave,Gao,Longhi}. Furthermore, spatially selective pumping has facilitated the excitation of Hermitian flat band \cite{Florent} and enabled the tuning of its localization length \cite{Ge_selective5}. Building upon the non-Hermitian extension of particle-hole symmetry, spatially selective pumping has been utilized to propose non-Hermitian flatbands \cite{NHFlatband_PRL,NHFlatband_PRJ} and zero-mode lasers with tunable spatial profiles \cite{zeromodeLaser}. Moreover, it provides a convenient route to study topologically protected photonic edge and corner states, without the need for resonant excitation. By focusing the pump at the edges of the system, such an approach has demonstrated localized edge states in one-dimensional (1D) lattices \cite{Poli,St-Jean,Zhao} and propagating chiral edge states in two-dimensional (2D) systems \cite{Bandres}. Notably, spatially selective pumping has shown the capability to redefine the system boundary, allowing steering of these chiral edge states on demand \cite{steering}.   

In this work, we probe a category of uniquely non-Hermitian topological edge and corner states with spatially selective pumping, i.e., those arising from the phenomenon known as the non-Hermitian skin effect \cite{Hatano,Longhi_gauge,Li_gauge,Song_gauge,Szameit_gauge,Zhang_gauge,Wang_Nature}. In the presence of a spatially uniform non-Hermitian gauge field, a significant proportion of the system's eigenmodes are localized toward one edge or corner of the system, which holds in both 1D and higher-dimensional systems. By varying this imaginary gauge field spatially, one can also achieve localization at any target position \cite{Ge_gauge}, including one corner, all corners, or any interior point, as recently demonstrated using a 2D array of optical micro-ring resonators \cite{Feng_gauge,Feng_gauge2,Feng_gauge3}. Despite their similarly localized spatial profiles to those rooted in Hermitian symmetry and topology \cite{Poli,St-Jean,Bandres}, spatially selective pumping has not been studied systematically for these non-Hermitian gauged laser arrays. 

Intriguingly, our study reveals an unexpected threshold behavior when selectively pumping either the head or the tail of the non-Hermitian skin modes: both cases exhibit the same lasing threshold and hence defy the conventional principle of spatially selective pumping. To shed light on this enigma, we uncover a previously overlooked phenomenon that provides a key insight into its understanding, i.e., energy exchange at non-Hermitian coupling junctions with the photonic environment, which does not occur with uniform gain or loss. With selective pumping, however, this energy exchange is nonzero and varies across the array, depending not only on the non-Hermitian gauge field but also the position of the pump. 

Importantly, we show that the usual non-Hermitian tight-binding model, while mathematically rigorous, provides an incomplete and even misleading physical explanation for this unusual threshold behavior. 
Utilizing a transfer matrix approach instead, we elucidate the mechanism of this anomalous threshold behavior, which is determined by the specific physical realization of the non-Hermitian gauge field (i.e., using gain, loss, or their mixture). Finally, we derive a generalized principle of selective pumping in non-Hermitian arrays, which shows that the decisive spatial overlap is given by the triple product of the pump, the lasing mode, and its biorthogonal partner.

\section{Conventional principle of selective pumping}

In order to elucidate the breakdown of the conventional principle of selective pumping (with ``spatially'' dropped for conciseness) in a non-Hermitian gauged array, below we provide a brief overview of this principle. Suppose $H$ is the passive tight-binding Hamiltonian of an array in the position basis $\{|n\rangle\}\,(n=1,2,\ldots,N)$, where $N$ is the size of the array. With the introduction of a pump with strength $D>0$ and a profile given by $F=\sum_n f_n|n\rangle \langle n|\,(f_n\geq0)$, the now active system can be described by $H' = (H+iDF)$ until the system reaches its lasing threshold, beyond which nonlinear effects such as gain saturation become important.

Selective pumping aims to increase the spatial overlap between the pump profile $F$ and a target mode $\Psi_\mu$, and when $\Psi_\mu$'s pump utilization surpasses other competing modes, this target mode becomes the first to reach its lasing threshold. More specifically, let $D_0,D_s$ be the thresholds of a passive mode $\Psi_0$ with uniform and selective pumping, respectively. It can be shown that (see Appendix \ref{sec:TH})
\be
D_s \approx \frac{D_0}{\Psi^T_0F\Psi_0}\label{eq:TH}
\ee
with the normalizations $\Tr{F}=N$ and $\Psi^T_0\Psi_0=1$. The denominator on the right hand side quantifies the spatial overlap between $\Psi_0$ and the pump profile, and the superscript ``$T$'' denotes the matrix transpose. A strong (weak) overlap then leads to a low (high) threshold with selective pumping.

\begin{figure}[b]
\includegraphics[clip,width=\linewidth]{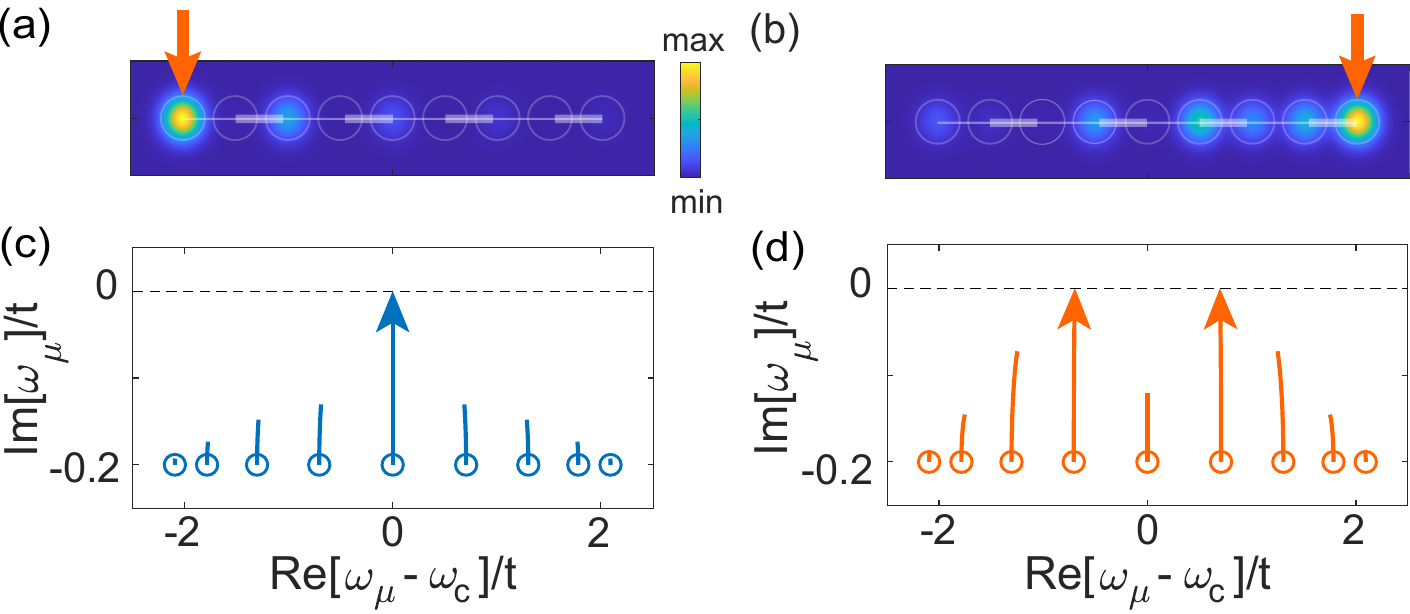}
\caption{\textbf{Selective pumping an SSH laser array}. (a,b) Intensity profile of the lasing mode at threshold when pumping the left and right cavity (orange arrows), respectively. A Gaussian peak is superposed in each cavity. (c,d) Corresponding trajectories of the resonances when increasing the pump. Line(s) with an arrow show the lasing mode(s), and open dots mark the passive resonances. Here $t'/t=1.2$, $\kappa_0/t=0.2$, and $N=9$. 
} \label{fig:ssh}
\end{figure}

Take the Su-Schrieffer-Heeger (SSH) array \cite{SSH}, for example: 
\be
H  = \sum_n (\omega_c-i\kappa_0) |n\rangle \langle n|  + \left(t_n|n+1\rangle \langle n|  + h.c.\right).\nonumber
\ee 
Here $\omega_c$ is the single-cavity frequency and $\kappa_0$ is the cavity loss. The nearest-neighbor (NN) coupling $t_n\in\mathbb{R}$ is given by $t$ ($t'$) when $n$ is odd (even), and $h.c.$ stands for the Hermitian conjugation of the first term in the brackets. In the passive case, i.e., without pump, all the energy eigenvalues of $H$ have the same imaginary part given by $-i\kappa_0$ [Fig.~\ref{fig:ssh}(c)]. 
With an odd number of cavities, one of them is a zero-mode featuring $\re{\omega_\mu}-\omega_c=0$ and an example of topological edge states, localized at the left boundary when $t'>t$ [Fig.~\ref{fig:ssh}(a)].
To excite this edge mode, we pump just the leftmost cavity to induce a strong overlap \cite{Poli,St-Jean}, and indeed, this mode is the first to reach its lasing threshold [Fig.~\ref{fig:ssh}(c)]. The approximation (\ref{eq:TH}) captures this behavior nicely: it gives $D_s N=2.74\kappa_0$, close to its actual value $D_s N=2.56\kappa_0$. When we pump the rightmost cavity instead [Fig.~\ref{fig:ssh}(b)], this zero-mode has a much reduced pump utilization, and a pair of band-edge modes that overlap better with the pump become the first lasing modes instead [Fig.~\ref{fig:ssh}(d)], at a higher lasing threshold $D_s N=4.36\kappa_0$. 

\section{Non-Hermitian gauged array}

\subsection{Anomalous threshold behavior}

\begin{figure}[h]
\includegraphics[clip,width=\linewidth]{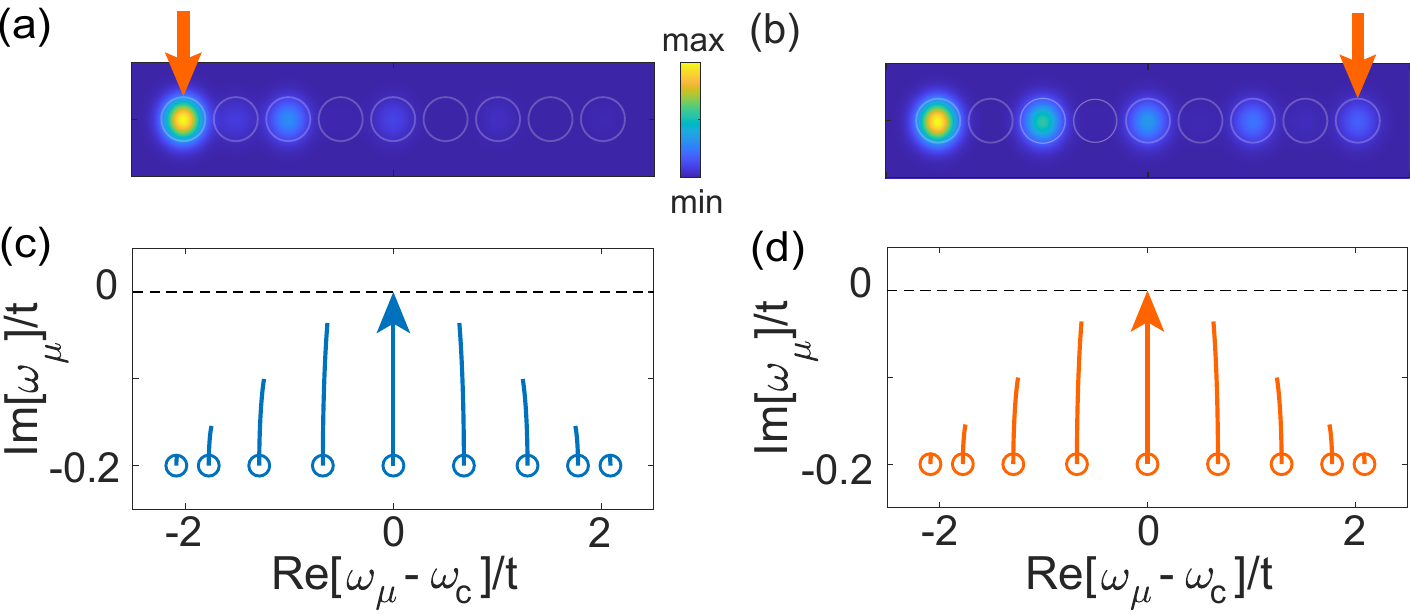}
\caption{\textbf{Selectively pumping a non-Hermitian gauged laser array}. The panels and parameters are the same as Fig.~\ref{fig:ssh}. The spatial profile of the zero-mode in panel (b) has a longer tail compared to that in panel (a). 
} \label{fig:gauged}
\end{figure}

Now let us focus on a non-Hermitian gauged laser array, featuring a stronger coupling from right to left ($t'>t$):
\be
H  = \sum_n (\omega_c-i\kappa_0) |n\rangle \langle n|  + \left(\,t|n+1\rangle \langle n|  + t'|n\rangle \langle n+1|\,\right). \nonumber
\ee 
The asymmetric couplings create a non-Hermitian gauge field \cite{Hatano} that localizes all modes toward the left edge. The zero-mode (also a non-Hermitian skin mode) has the strongest localization [Fig.~\ref{fig:gauged}(a)], which can be quantified using the inverse participation ratio (i.e., $\text{IPR}={(\sum_i |\Psi_i|^2 )^2}/{\sum_i |\Psi_i|^4}\approx4.0$ with $t'/t=1.2$ and $N=9$). 

When we pump the leftmost cavity, the zero-mode overlaps strongly with the pump and is the first lasing mode as expected [Fig.~\ref{fig:gauged}(c)]. With the same $t,t'$ as in the SSH array, this zero-mode has an IPR identical to that of the topological edge state in the SSH array. According to Eq.~(\ref{eq:TH}), we then expect that they should have similar if not identical lasing thresholds. However, we find $D_s N=4.22\kappa_0$ here instead, more than 50\% higher than the SSH array. Probably even more surprisingly, we find that this zero-mode localized on the left edge is again the first lasing mode when we pump the rightmost cavity [Fig.~\ref{fig:gauged}(b)], with exactly the same threshold [Fig.~\ref{fig:gauged}(d)]. This unexpected finding thus defies the conventional principle of selective pumping represented by Eq.~(\ref{eq:TH}).  

\subsection{Symmetry arguments}

To understand this unusual threshold behavior, we first note that the two configurations, i.e., pumping the leftmost or the rightmost cavities (denoted by Configurations A, B with pump profile $F_A,F_B$ and threshold $D_s^{(A)},D_s^{(B)}$), can be mapped into each other.
More specifically, the left-right mirror reflection $P$ (an anti-diagonal matrix with 1's) maps the left pump in Configuration A to the right pump in Configuration B, i.e., $PF_AP^{-1}=F_B$. It also exchanges the couplings $t,t'$, which can be restored by the imaginary gauge transformation given by the diagonal matrix $G=\text{diag}[1,t/t',(t/t')^2,\ldots]$. Because the latter does not change the onsite potential \cite{ge_gauge}, here given by just $(\omega_c-i\kappa_0)$ in $H$, we identify $S\equiv GP$ as a symmetry of $H$ (i.e., $SHS^{-1}=H$). Similarly, we find $G F_B G^{-1}=F_B$ using the property that $F_B,G$ are both diagonal matrices, and hence 
\be
S F_A S^{-1} = G (PF_AP^{-1}) G^{-1}=G F_B G^{-1}=F_B.
\ee
Altogether, we derive 
\be
\hspace{-2mm} S(H+iD^{(A)}_sF_A)S^{-1} = H+iD^{(A)}_sF_B \equiv H'_B. \label{eq:transformation} 
\ee
This expression then indicates $H'_B$, in Configuration B, has the same eigenvalues as $H+iD^{(A)}_sF_A$. The latter represents the system at threshold in Configuration A, with the zero-mode on the real axis and all other modes in the lower half of the complex plane [Fig.~\ref{fig:gauged}(c)]. Therefore, $H'_B$ also represents the system at threshold but in Configuration B, i.e., $D_s^{(B)} = D_s^{(A)}$, with the zero-mode again being the lasing mode [Fig.~\ref{fig:gauged}(d)]. We note, though, the wave functions $\Psi^{(A)}_s,\Psi^{(B)}_s$ of the zero-mode at threshold in these two configurations are different, i.e., $\Psi^{(B)}_s= S\Psi^{(A)}_s\neq \Psi^{(A)}_s$; they are both localized at the left edge but with different tail lengths [Figs.~\ref{fig:gauged}(a,b)]. 

\subsection{Energy exchange with the environment at non-Hermitian coupling junctions}

Next, to understand the breakdown of the conventional principle of selective pumping, we analyze the dynamical equation for the intensity in each cavity:
\be
\frac{d|\psi_n|^2}{dt} = 2(DF_{nn}-\kappa_0)|\psi_n|^2 + {\cal J}_{n,n+1} + {\cal J}_{n,n-1},\label{eq:dynamics}
\ee
where $F_{nn}$ is the pump strength in the $n$th cavity and 
\be
{\cal J}_{n,n+1} = it'^*\psi^*_{n+1} \psi_{n} + c.c.,\,\, {\cal J}_{n,n-1} = it^*\psi^*_{n-1} \psi_{n} + c.c. \nonumber
\ee
are the inter-cavity power flows from cavity $n+1$ to $n$ and from cavity $n-1$ to $n$ \cite{Ge_PRA_2017b}, respectively. Here we have used the notation $[\psi_1,\psi_2,\ldots]^T$ for the lasing mode at threshold, and $c.c.$ stands for the complex conjugation of the first term. 

Equation~(\ref{eq:dynamics}) indicates that there are two power sources or drains for each cavity, i.e., the onsite term given by $P_n = 2(DF_{nn}-\kappa_0)|\psi_n|^2$, and the exchange term $p_n = {\cal J}_{n,n+1} + {\cal J}_{n,n-1}$ with its neighbors via the coupling junctions \cite{bibnote:2}. At the laser threshold, a self-sustained oscillation means that the left hand side of Eq.~(\ref{eq:dynamics}) becomes zero for each cavity, and hence we have the following power relation:
\be
\sum_n P_n =-\sum_n p_n = -\sum_{n=1}^{N-1} g_{(n,n+1)}.\label{eq:sum}
\ee
In the last step we have rearranged the summation using $g_{(n,n+1)} \equiv {\cal J}_{n,n+1} + {\cal J}_{n+1,n}$, which is the power gained (if positive) or dissipated (if negative) at the coupling junction between cavities $n$ and $n+1$. 

In a system with symmetric couplings (i.e., $t'=t^*$), it is clear that ${\cal J}_{n,n+1}=-{\cal J}_{n+1,n}$ from their definitions, indicating all $g_{(n,n+1)}=0$, i.e., power flows from one cavity to its neighbors without being amplified or dissipated. Therefore, gain and loss for the entire lattice only comes from the onsite terms, whether or not the pump is uniform. Equation~(\ref{eq:sum}) then gives $D_0 = \kappa_0$ with uniform pumping and $D_s=\kappa_0/(\Psi_s^\dagger F\Psi_s)$ with selective pumping. $\Psi_s$ is the lasing mode at threshold with selective pumping and normalized by $\Psi_s^\dagger\Psi_s=1$. We then recover Eq.~(\ref{eq:TH}) if $\Psi_0\approx\Psi_s\in\mathbb{R}$.

This derivation shows clearly that the conventional principle of selective pumping is based on the assumption that the system does not exchange power with the photonic environment via the inter-cavity couplings (at least not strongly). Note that this condition \textit{holds} even in our non-Hermitian gauged array at threshold, \textit{if} the pump is uniform: with real couplings $t$ and $t'$, $H'=H+iD_0F$ (at threshold) is real-valued and has a real, non-degenerate spectrum. Therefore, we find
\be
H'\Psi_0^*=(H'\Psi_0)^*=(\omega^\text{(TH)}\Psi_0)^*=\omega^\text{(TH)}\Psi_0^*
\ee
with $H'\Psi_0=\omega^\text{(TH)}\Psi_0$ (also see Appendix~\ref{sec:TH}). The observation then indicates that $\Psi_0^*=\Psi_0$ is real, 
which renders all ${\cal J}_{n,n+1},{\cal J}_{n+1,n}$ (and hence $g_{(n,n+1)}$) zero.

\begin{figure}[b]
\includegraphics[clip,width=\linewidth]{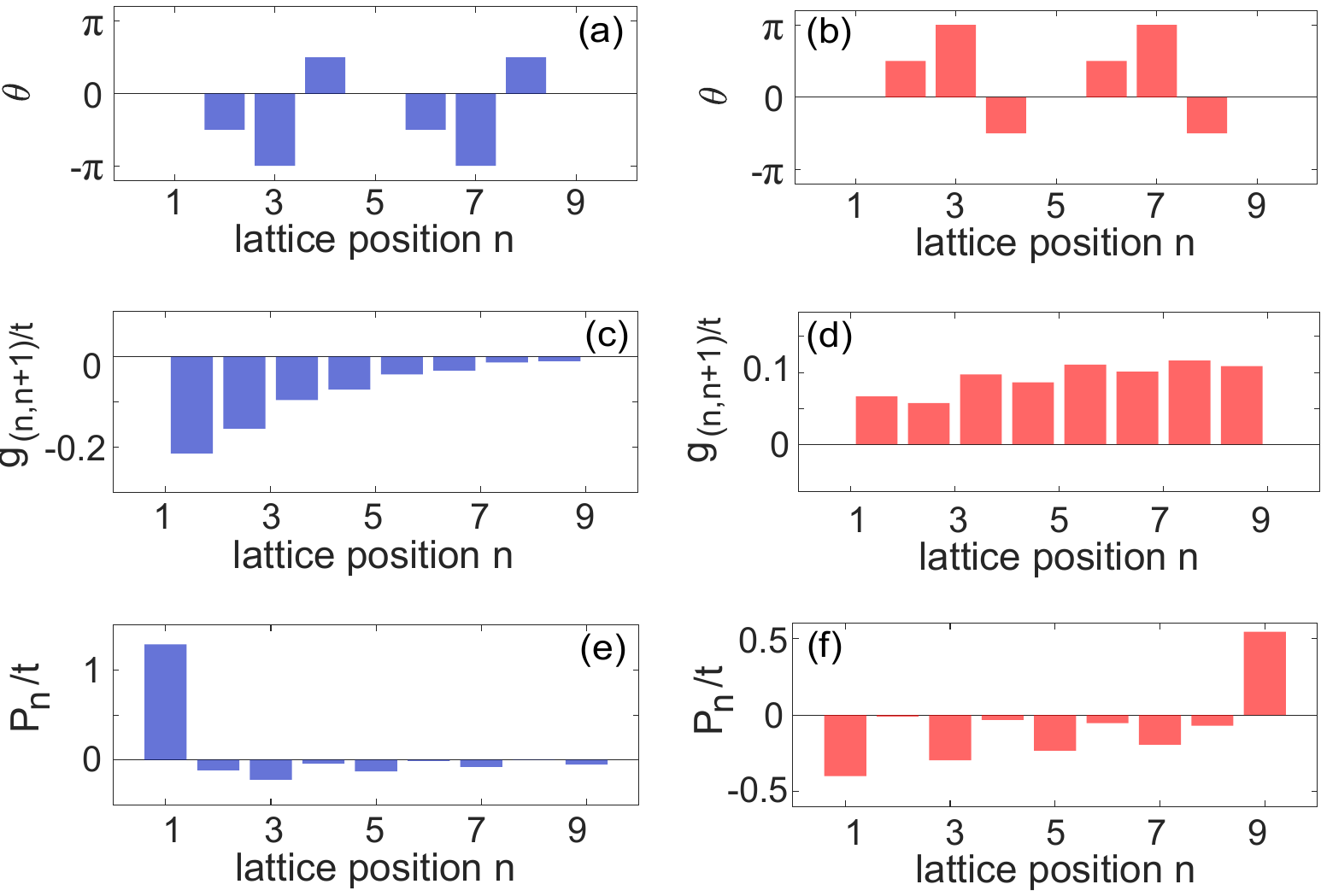}
\caption{\textbf{Properties of the zero-mode at threshold in a non-Hermitian gauged laser array.} (a,b) Phase of its wave function in Configurations A and B. (c,d) Power dissipated ($g_{(n,n+1)}<0$) and gained ($g_{(n,n+1)}>0$) at each coupling junction in these two configurations, plotted at half integer positions and with the normalization $\psi_1=1$. (e,f) The corresponding power gained ($P_n>0$) or dissipated ($P_n<0$) onsite. 
} \label{fig:flow}
\end{figure}

With selective pumping, however, these eigenstates become complex-valued in our non-Hermitian gauged array. For the lasing zero-mode in particular, its wave function has a staggering phase in both Configurations A and B [Figs.~\ref{fig:flow}(a,b)], i.e., with a $\pm\pi/2$ relative phase between neighboring cavities due to non-Hermitian particle-hole symmetry \cite{zeromodeLaser}. As a result, $g_{(n,n+1)}=[({t^*}/{t'})-1]{\cal J}_{n,n+1}$ is no longer zero due to the asymmetric couplings, and hence the system exchanges power with its photonic environment via the coupling junctions, leading to the breakdown of the conventional principle of selective pumping. 


\subsection{Generalized principle of selective pumping}

Nevertheless, a generalized principle of selective pumping can be established (see Appendix \ref{sec:TH}): 
\be
D_s\approx\frac{D_0}{\tilde{\Psi}^T_0 F \Psi_0}, \label{eq:TH2}
\ee
where $\tilde{\Psi}_0^T$ is the corresponding left eigenstate of the passive $H$ (i.e., the biorthogonal partner of $\Psi_0$), normalized by $\tilde{\Psi}^T_0 \Psi_0=1$ \cite{bibnote:1}. It is then clear that the decisive spatial overlap now that determines the lasing threshold is not just between the pump profile and the lasing mode, but also with its biorthogonal partner $\tilde{\Psi}_0^T$. When $H$ is symmetric as in the SSH array, we find $\tilde{\Psi}_0=\Psi_0$ and recover Eq.~(\ref{eq:TH}).

Equation~(\ref{eq:TH2}), though just an approximation, further corroborates our findings of the identical thresholds in Configurations A and B: $\tilde{\Psi}_0$ (without the transpose) is a right eigenstate of $H^T$, which is mapped to $H$ by the aforementioned mirror reflection $P$. Therefore, $\tilde{\Psi}_0$ and ${\Psi}_0$ are mirror-symmetric partners, and any two pump profiles that are also mirror-symmetric partners give the same triple product $\tilde{\Psi}^T_0 F \Psi_0$, and in turn, the same lasing threshold according to Eq.~(\ref{eq:TH2}). We find $D_s=5\kappa_0$ for \textit{both} Configurations A and B using Eq.~(\ref{eq:TH2}), which agrees with its numerical value ($4.22\kappa_0$) qualitatively. 

\subsection{Failure of the tight-binding model}

Following our previous discussion of the power relations in the non-Hermitian gauged array, next we offer a physical understanding of this identical threshold in these two configurations. We find all $g_{(n,n+1)}$'s are negative in Configuration A [Fig.~\ref{fig:flow}(c)], indicating that the system dissipates energy into the environment at every coupling junction. The situation is reversed in Configuration B [Fig.~\ref{fig:flow}(d)], where the system receives power from the environment at all coupling junctions. 
This contrast provides an explanation of their identical lasing threshold: due to the vastly different overlaps between these two pump configurations and the non-Hermitian skin mode, the system receives a net onsite gain $\sum_n P_n = 3.18\kappa_0$ in Configuration A [Fig.~\ref{fig:flow}(e)] but a net onsite loss $-\sum_n P_n = 3.73\kappa_0$ in Configuration B at threshold [Fig.~\ref{fig:flow}(f)]. Nevertheless, they are compensated by dissipating and gaining exactly the same amounts via the coupling junctions.   

\begin{figure}[b]
\includegraphics[clip,width=\linewidth]{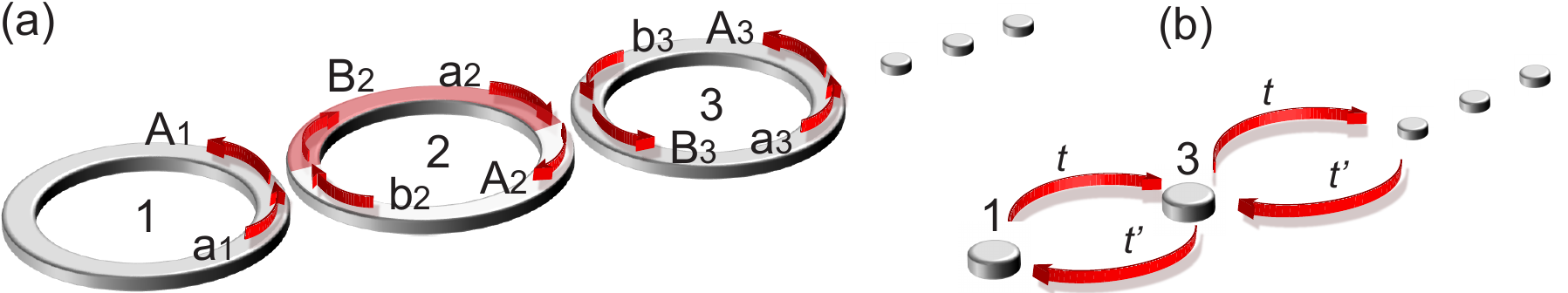}
\caption{\textbf{Schematics of a non-Hermitian gauged array.} (a) Couplings of CCW modes in cavity rings (1, 3, $\ldots$) via auxiliary rings (2, $\ldots$) with two halves of different Im$[n]$'s in the transfer matrix analysis. (b) The corresponding tight-binding model.} \label{fig:couplings}
\end{figure}

A closer examination of this seemingly satisfactory explanation suggests, however, either it is misleading in certain non-Hermitian gauged arrays or it excludes these systems from exhibiting the aforementioned anomalous threshold behavior: one common approach to realize asymmetric couplings and the resulting non-Hermitian gauge field in photonic systems is using auxiliary rings with different gain and/or loss halves as couplers (see, for example,  cavity 2 in Fig.~\ref{fig:couplings}; also Refs.~\cite{Feng_gauge,Feng_gauge2,Feng_gauge3}). In the case that the auxiliary rings are passive with lossy and lossier halves, they simply \textit{cannot} provide the gain needed in Configuration B at the coupling junctions. Similarly, if the auxiliary rings are active with gain and more gain in the two halves, they \textit{cannot} induce the loss needed in Configuration A.

These observations highlight a significant drawback in relying on the tight-binding model to describe the physical system with asymmetric couplings: it does not provide information on the nature of each coupling, which is essential for understanding the energy exchange between the system and its environment. Furthermore, a potentially graver concern arises when the supermodes formed by the couplings of the auxiliary rings may even lase before the non-Hermitian skin modes. To tackle these issues, below we analyze the non-Hermitian gauged array using a transfer matrix approach, from which we gain a more accurate and detailed understanding of the aforementioned anomalous threshold behavior.


\subsection{Transfer matrix analysis}

We consider the couplings of counterclockwise (CCW) modes in the odd-numbered rings shown in Fig.~\ref{fig:couplings}(a), which are the cavities considered in the tight-binding model. We define $e_p = e^{in_pkL}\,(p=1,3,\ldots,N)$ where $n_p$ is the refractive index in the $n$th ring and $k$ is the free-space wave vector. These rings have the same $\re{n_p}$ and length $L$ but can have different $\im{n_p}$'s depending on the pump profile. The amplitudes of the CCW mode in ring 1 and those of the clockwise (CW) mode in ring 2 (one of the auxiliary rings) satisfy \cite{Yariv,inverseVernier}
\be
\begin{pmatrix}
A_1\\
B_2
\end{pmatrix} =
S
\begin{pmatrix}
a_1\\
b_2
\end{pmatrix},\quad
S =
\begin{pmatrix}
s & iJ \\
iJ^* & s^*
\end{pmatrix}.
\ee
$S$ is the unitary scattering matrix satisfying $|s|^2+|J|^2=1$. To propagate these amplitudes down the array, we rewrite this equation as
\be
\begin{pmatrix}
A_1\\
a_1
\end{pmatrix} =
M_c
\begin{pmatrix}
B_2\\
b_2
\end{pmatrix},\quad
M_c = M_c^{-1} = \frac{1}{iJ}
\begin{pmatrix}
s & -1\\
1 & -s
\end{pmatrix}.
\ee
Next, the propagation of waves inside cavity 2 can be written as
\be
\begin{pmatrix}
B_2\\
b_2
\end{pmatrix} =
M_2
\begin{pmatrix}
A_2\\
a_2
\end{pmatrix},\quad
M_2 = e_u^{-1}
\begin{pmatrix}
0 & 1\\
e_de_u & 0
\end{pmatrix} \label{eq:M2}
\ee
with $e_u=e^{in_ukL_u}$ and $e_d=e^{in_dkL_d}$. Here $n_u,L_u$ ($n_d,L_d$) are the refractive index and length of the upper (lower) half of the auxiliary ring. To achieve $t'>t$ in the non-Hermitian gauged array as in Fig.~\ref{fig:gauged}, we then require $\im{n_d}<\im{n_u}$, which represent gain (loss) when negative (positive). 

By repeating the same procedure for all $N$ ring resonators in the array, we find
\be
\begin{pmatrix}
A_1\\
a_1
\end{pmatrix}
\equiv
M
\begin{pmatrix}
B_N\\
b_N
\end{pmatrix},\quad
M \equiv
(\Pi_{p=2}^{N-1} M_cM_{p})M_c, \label{eq:M}
\ee
where the transfer matrices inside the cavity rings are
\be
M_p = 
\begin{pmatrix}
0 & e_p^{-1/2}\\
e_p^{1/2} & 0
\end{pmatrix}\,\,(p=3,5,\ldots,N-2).\nonumber 
\ee
Assuming all auxiliary rings are identical, we also have $M_2=M_4=\ldots=M_{N-1}$.

\begin{figure}[t]
\includegraphics[clip,width=\linewidth]{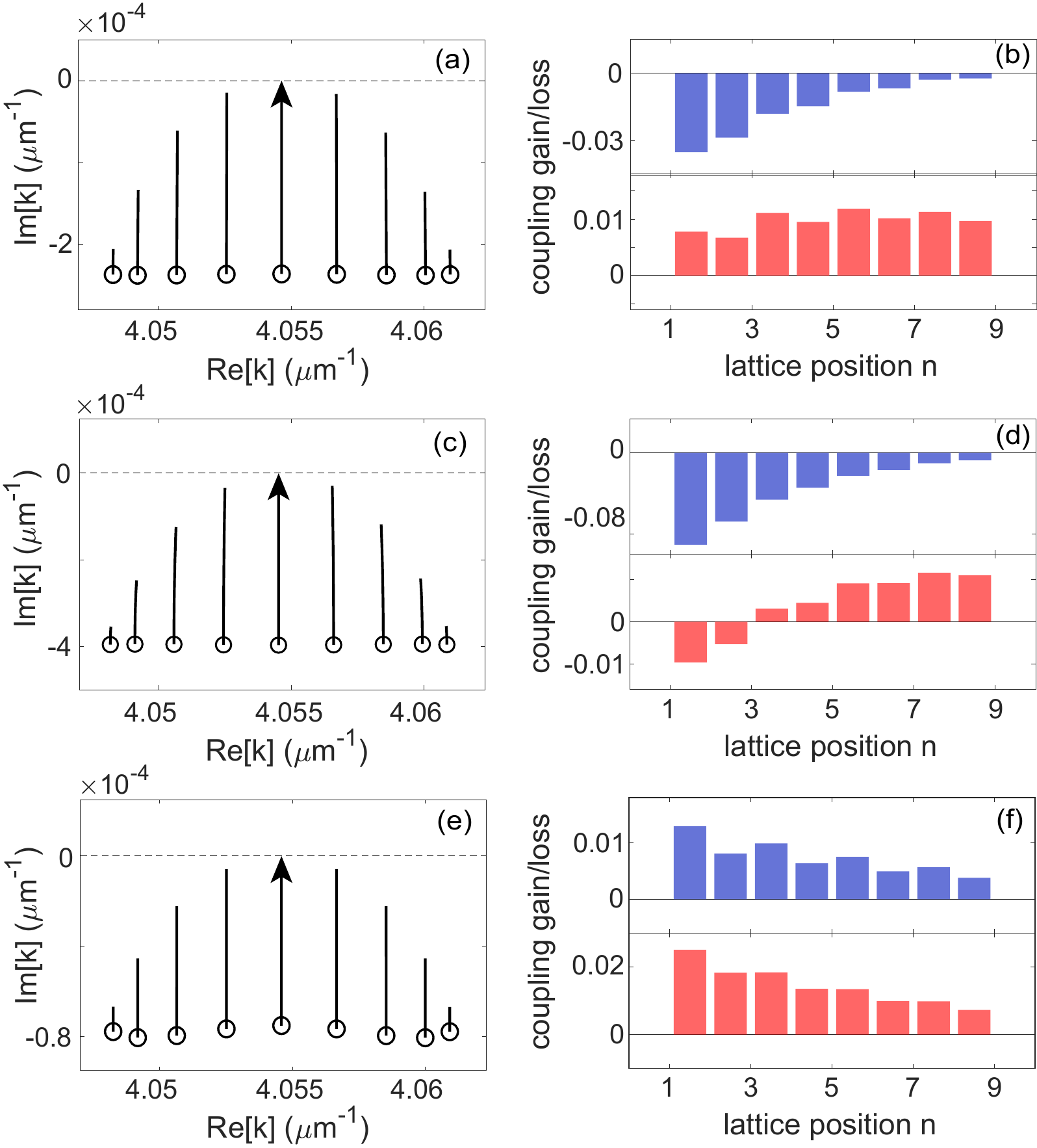}
\caption{\textbf{Transfer matrix analysis of non-Hermitian gauged arrays.} (a) Trajectories of the resonances when pumping either the first or the last cavity ring. Symbols are the same as in Fig.~\ref{fig:gauged}(c). (b) Power gained (positive) or dissipated (negative) at coupling junctions at threshold. Upper (lower) panel is for Configuration A (B) with the normalization $A_1=1$ in the first cavity. Parameters used are $r=4.75\,\mu{m}$ and $r'=4.70\,\mu{m}$ in the cavity rings and auxiliary rings respectively, with $\re{n}=2.7$ in all rings. Passive cavity rings have $\im{n}=2\times10^{-4}$. $s=0.7$, $J=\sqrt{1-s^2}$, and $\im{n_u} = 1.5\times10^{-3} = -\im{n_d}$. (c,d) Same as (a,b) but with $\im{n_u} = 2\times10^{-3} = -2\im{n_d}$; (e,f) Same as (a,b) but with $\im{n_u} = 10^{-3} = -\im{n_d}/2$.  
} \label{fig:Tmat}
\end{figure}

The resonances of the array are then found by combining Eq.~(\ref{eq:M}) with the propagation equations in the first and last (cavity) rings, i.e., $a_1=A_1e_1$ and $b_N=B_Ne_N$, which leads to
\be
e_1 = \frac{M_{21}+M_{22}e_N}{M_{11}+M_{12}e_N}. \label{eq:Nrings}
\ee
$M_{ij}$'s are the matrix elements of the total transfer matrix $M$, and they are proportional to $e_u^{-N_a}$ [see the expression for $M_2$ in Eq.~(\ref{eq:M2})], where $N_a=(N-1)/2$ is the number of auxiliary rings. This factor is cancelled in the denominator and numerator on the right hand side of Eq.~(\ref{eq:Nrings}), whose only dependence on $e_u,e_d$ comes from the product $e_ue_d$ in $M_p\,(p=2,4,\ldots)$. 
This observation shows explicitly the non-Hermitian gauge symmetry: changing $n_uL_u$, $n_dL_d$ (and in particular their imaginary parts) does not affect the resonance frequencies, as long as $\sigma \equiv n_uL_u+n_dL_d$ is a constant. Therefore, we recover the anomalous threshold behavior shown in Fig.~\ref{fig:gauged}, i.e., pumping the head or the tail of a non-Hermitian skin mode leads to the same lasing threshold. 

Next, we show that the mechanism of this anomalous threshold behavior depends on the specific physical realization of the non-Hermitian gauge field. Figure~\ref{fig:Tmat} shows the results for a non-Hermitian gauged array formed by nine cavity rings and eight auxiliary rings. Three cases are presented using auxiliary rings with balanced gain and loss, net loss, and net gain, respectively. They feature the same imaginary gauge field (see Appendix \ref{sec:Tmat})
\be
{t'}/{t} \approx e^{\im{n_uL_u-n_dL_d}k_0}=1.2 \label{eq:iGuage}
\ee
as in the tight-binding model shown in Fig.~\ref{fig:gauged}, where $k_0$ is the free-space wave vector of the zero-mode. Below we define the coupling gain/loss via the auxiliary ring $p$ by $(|a_p|^2-|B_p|^2)+(|b_p|^2-|A_p|^2)$, with the two intensity differences occurring at the upper and lower halves, respectively [see, for example, the one labeled by $p=2$ in Fig.~\ref{fig:couplings}(a)]. For convenience, both the wave amplitudes and the coupling gain/loss here are dimensionless, scaled by their nature units. 

The first case exhibits similar results to the tight-binding model, i.e., with loss (gain) at all coupling junctions in Configuration A (B) [Fig.~\ref{fig:Tmat}(b)]. 
The threshold 
in Configurations A and B is found at $\im{n}=-7.96\times10^{-4}$, 
and the trajectories of the non-Hermitian skin modes [Fig.~\ref{fig:Tmat}(a)] are similar to those in Fig.~\ref{fig:gauged} obtained from the tight-binding model. While the auxiliary-ring resonances have lower losses in the passive case, they do not reach their threshold when the zero-mode starts lasing with selective pumping (see Fig.~\ref{fig:17ring_SM} in Appendix \ref{sec:Tmat}). 

If we assume a stronger cavity loss (e.g., $\im{n}=2\times10^{-3}$), the spatial profile of the lasing mode in either Configuration A or B is determined mainly by the localized pump, i.e., with a peak at the pump position (not shown). This is the regime of zero-mode lasing with tunable spatial profiles \cite{zeromodeLaser}, and these two configurations still share the same lasing threshold. The same situation takes place when we use only loss or gain in the auxiliary rings to achieve the same imaginary gauge field (e.g., $\im{n_d}=4\im{n_u}=4\times10^{-3}$) (not shown). 

In order for the first lasing mode at threshold to be a non-Hermitian skin mode peaked at the leftmost cavity in both Configurations A and B, below we consider auxiliary rings with net loss and gain instead. By increasing (decreasing) the loss (gain) in the upper (lower) halves of the auxiliary rings by $\im{\Delta n}=5\times10^{-4}$, the second case in Fig.~\ref{fig:Tmat} shows results that differ from the tight-binding model: while all coupling junctions still dissipate power in Configuration A, not all of them gain power in Configuration B [Fig.~\ref{fig:Tmat}(d)]. 
The largest deviation from the tight-binding model is the last case shown in Figs.~\ref{fig:Tmat}(e-f), where we achieve the same asymmetric couplings (and imaginary gauge field) as in the cases above by increasing (decreasing) the gain (loss) in the lower (upper) halves of the auxiliary rings by $\im{\Delta n}=-5\times10^{-4}$ instead. Now the system gains power from all auxiliary rings in both Configurations A and B, and their identical lasing threshold is understandably lower at $\im{n}=-1.26\times10^{-4}$, thanks to the enhanced gain in the auxiliary rings. 

The passive resonances in Figs.~\ref{fig:Tmat}(a,c,e) have almost the same spacing, which indicates that the average coupling $\tilde{t}\equiv tt'$ is roughly equal in these three cases, besides the imaginary gauge field $t'/t$. This observation then shows that both $t$ and $t'$ are fixed in these cases approximately,  despite that we have changed the gain and loss in the auxiliary rings significantly. 

This finding first suggests that the coupling $t$ ($t'$) is not simply determined by the upper (lower) halves of the auxiliary rings, as one may have anticipated from comparing the two schematics in Fig.~\ref{fig:couplings} or assumed in previous studies. Furthermore, it also shows that different realizations of the same non-Hermitian gauge field demand contrasting power exchanges with the photonic environment to produce the same threshold in Configurations A and B, which is beyond the analyzing capability of the tight-binding model. This is true even after we take into account the different effective cavity decays of the passive resonances shown in Figs.~\ref{fig:Tmat}(a,c,e) (see Appendix \ref{sec:decay}). 

As we have hinted in the motivations of adopting the transfer matrix analysis, here another crucial drawback of the tight-binding model is its inability to describe the supermodes formed by the auxiliary ring resonances, which have much lower thresholds than the non-Hermitian skin modes in the third case above; they are already lasing before we even pump the cavity rings (see Fig.~\ref{fig:17ring_SM} in Appendix \ref{sec:Tmat}).

We also mention in passing that for the CW modes in the cavity rings (instead of CCW modes considered above), their couplings lead to the same resonances in the transfer matrix analysis: we just need to switch $e_u$ and $e_d$ in the analysis above, which leads to the same equation (\ref{eq:Nrings}) thanks to its non-Hermitian gauge symmetry. Consequently, degenerate CW and CCW non-Hermitian skin modes 
are localized on the opposite edges of a 1D array due to the opposite non-Hermitian gauge fields they experience. The anomalous threshold behavior discussed above then manifests itself as a lasing mode with a symmetric intensity profile at threshold, localized on both edges before nonlinear gain saturation and modal interaction play a role. We further note that the same anomalous threshold behavior takes place in higher dimensional systems \cite{Feng_gauge3} as well. For a square lattice with a uniform non-Hermitian gauge field in both the $x$ and $y$ directions, pumping either of the four corners (i.e., the head, tail, and two ``wings'') leads to the same lasing threshold (see Appendix \ref{sec:2D}). 

\section{Conclusion}

In summary, we have reported an anomalous threshold behavior when pumping either the head or tail of a non-Hermitian skin mode; these two configurations lead to the same lasing threshold and hence defy the conventional principle of selective pumping. We have given explanations to this behavior from both the mathematical and physical perspectives, by introducing a transfer matrix analysis that remedies two crucial drawbacks of the usual tight-binding model, i.e., the lack of information on the physical realization of the imaginary gauge field and whether the auxiliary rings lase first. Our study provides a glimpse into how the two forms of non-Hermiticity, i.e., asymmetric couplings and a complex onsite potential \cite{Rivero_PRL_2022}, interact in a synergetic fashion, which may stimulate further explorations of their collective effects in photonics and related fields. 

\begin{acknowledgments}

This project is supported by NSF under Grant Nos. PHY-1847240 and ECCS-1846766.
\end{acknowledgments}

\appendix

\section{Threshold analysis in the tight-binding model}
\label{sec:TH}

To quantify the effect of selective pumping in the tight-binding model of a laser array, we analyze the equation that determines the lowest lasing threshold with uniform pumping:
\be
(H+iD_0\bm{1})\Psi_0 = \omega^\text{(TH)}_0\Psi_0.
\ee
Here $H$ is the effective Hamiltonian of the passive system (i.e., with cavity decay $\kappa_0$ but no gain), 
$D_0$ represents the pump strength at the lasing threshold, and $\bm{1}$ is the identity matrix representing uniform pumping. $\omega_0^{\text{(TH)}}\in\mathbb{R}$ and $\Psi_0$ are the lasing frequency and wave function at threshold. Note that the eigenstates of $H+iD_0\bm{1}$ are the same as those of $H$, and its eigenvalues are merely shifted from those of $H$ (denoted by $\{\omega_\mu\}$) by $iD_0$. In other words, 
\be
\omega^\text{(TH)}_0 = \re{\omega_0},\quad D_0 = -\im{\omega_0}, \label{eq:TH_uniform}
\ee
where $\omega_0$ is the passive resonance that evolves into the lasing mode at threshold.

Below we refer to this basis (including the lasing mode at threshold) and the corresponding left eigenstates of $H$ by $\{\Psi_\mu\}$ and $\{\tilde{\Psi}^T_\mu\}$\,$(\mu=0,1,2,\ldots)$, and they satisfy the following biorthogonal relation after normalization:
\be
\tilde{\Psi}^T_\mu {\Psi}_\nu = \delta_{\mu\nu}.
\ee
Here we have assumed that none of these states are at an exceptional point, which would make $\tilde{\Psi}^T_\mu\Psi_\mu=0$ and the basis $\{\Psi_\mu\}$ incomplete. 

Now with selective pumping, we represent the pump profile $f(\vec{r})$ by a diagonal and positive semi-definite matrix $F$ normalized by $\Tr{F}=N$, where $N$ is the size of the lattice. We then have
\be
(H+iD_sF)\Psi_s = \omega^\text{(TH)}_s\Psi_s,\label{eq:selectivePumping}
\ee
where $D_s$ is the new threshold. $\omega^\text{(TH)}_s\in\mathbb{R}$ and $\Psi_s$ are the modified lasing frequency and wave function at threshold. Their formal solutions can be obtained by the expansion $\Psi_s = \sum_\mu a_\mu\Psi_\mu$, which leads to
\be
\sum_\mu a_\mu(\omega_\mu-\omega^\text{(TH)}_s + iD_sF)\Psi_\mu=0
\ee 
or 
\be
{\cal F}\bm{A} = \frac{1}{D_s}{\cal W}\bm{A}, \label{eq:SALT}
\ee
where ${\cal F}$ is a square matrix with elements ${\cal F}_{\mu\nu}=\tilde{\Psi}^T_\mu F \Psi_\nu$, $\bm{A} = [a_0,a_1,\ldots,a_N]^T$, and ${\cal W}$ is a diagonal matrix with elements ${\cal W}_{\mu\mu}=i(\omega_\mu-\omega^\text{(TH)}_s)$. 

Equation~(\ref{eq:SALT}) is a system of $N$ complex linear equations, and by fixing the normalization of $\bm{A}$ (e.g., requiring $a_0=1$), we can solve for the $N$ unknowns: the $(N-1)$ amplitudes and the $(N-1)$ phases of $\{a_{\mu>0}\}$, the lasing frequency $\omega^\text{(TH)}_s$, and the threshold $D_s$. One strategy to find all solutions of Eq.~(\ref{eq:SALT}) is treating it as a parametrized eigenvalue problem \cite{SALT_Science,SALT_Nonlinearity,Ge_SciRep}: 
\be
{\cal F}\bm{A}_\mu = \lambda_\mu{\cal W}\bm{A}_\mu.
\ee
By tuning the real frequency $\omega^\text{(TH)}_s$ in $\cal W$, we monitor the generalized eigenvalues $\lambda_\mu$'s in the complex plane. When one of them crosses the real axis, the inverse of that real eigenvalue gives one (linear) lasing threshold. The corresponding eigenstate of $\bm{A}_\mu$ is the wave function of that lasing mode at threshold in the basis $\{\Psi_\mu\}$, and the tuning parameter $\omega^\text{(TH)}_s$ at which this crossing happens is the corresponding lasing frequency. 

If $|{\cal F}_{00} a_0|$ is much greater than  $|\sum_{\mu>0}{\cal F}_{0\mu} a_{\mu}|$, which usually takes place in a lattice with strong couplings and weak losses (i.e., with modes having high quality factor), the first row of the matrix equation in (\ref{eq:SALT}) gives
\be
D_s\approx\frac{i(\omega_0-\omega^\text{(TH)}_s)}{{\cal F}_{00}} = \frac{D_0}{\tilde{\Psi}^T_0 F \Psi_0} + o(\epsilon).\label{eq:TH_SM}
\ee
Here we have used the expressions for $D_0$ and $\omega^\text{(TH)}_0$ in Eq.~(\ref{eq:TH_uniform}), and we note $\epsilon\equiv\omega^\text{(TH)}_s-\omega^\text{(TH)}_0$ vanishes if a zero-mode lases with both uniform and selective pumping. 

To understand better the requirement on the lasing mode in this approximation, we note that it can also be derived using $\omega^\text{(TH)}_s\approx\omega^\text{(TH)}_0$ and $\Psi_s\approx\Psi_0$, i.e., the lasing frequency and the spatial profile of the lasing mode should stay roughly the same with selective pumping. We then find $(D_0\bm{1}-D_sF)\Psi_0 \approx 0$ and recover Eq.~(\ref{eq:TH_SM}). 

The denominator on the right hand side of Eq.~(\ref{eq:TH_SM}) represents the spatial overlap between the lasing mode, the pump profile, \textit{and} its biorthogonal partner (i.e., the corresponding left eigenstate of $H$). When $H$ is symmetric as in the SSH array, we find $\tilde{\Psi}_0=\Psi_0$, with which we derive Eq.~(\ref{eq:TH}) in the main text. In the continuous limit, we then recover the more conventional form of this denominator, i.e., $\int d\vec{r} f(\vec{r})\Psi^2_0(\vec{r})$ \cite{Ge_selective1}.

In the main text, we have used Eq.~(\ref{eq:TH_SM}) to further corroborate the identical threshold when pumping the leftmost and rightmost cavity in a non-Hermitian gauged array. We note though not all observations based on Eq.~(\ref{eq:TH_SM}) hold due to its approximative nature. For example, given the structure $\Psi_0\propto[1,0,(t'/t),0,\ldots,0,(t'/t)^{N-1}]^T$ and $\tilde{\Psi}_0\propto[(t'/t)^{N-1},0,(t'/t)^{N-2},0,\ldots,0,1]^T$ of the zero-mode in the non-Hermitian gauged array, we find $\tilde{\Psi}^T_0F\Psi_0 \propto \sum_{n\in\text{odd}}F_{nn}$ where $F_{nn}$ is the pump strength in the $n$th cavity. As a result, selective pumping any odd-numbered cavity gives the same threshold according to Eq.~(\ref{eq:TH_SM}), but the actually thresholds in these pump configurations differ. For example, the zero-mode has a threshold of $D_s N=4.81\kappa_0$ when pumping the middle cavity, and we have seen in the main text that this value is $4.49\kappa_0$ instead when pumping the leftmost or the rightmost cavity.

\section{Different cavity decays}
\label{sec:decay}

\begin{figure}[t]
\includegraphics[clip,width=\linewidth]{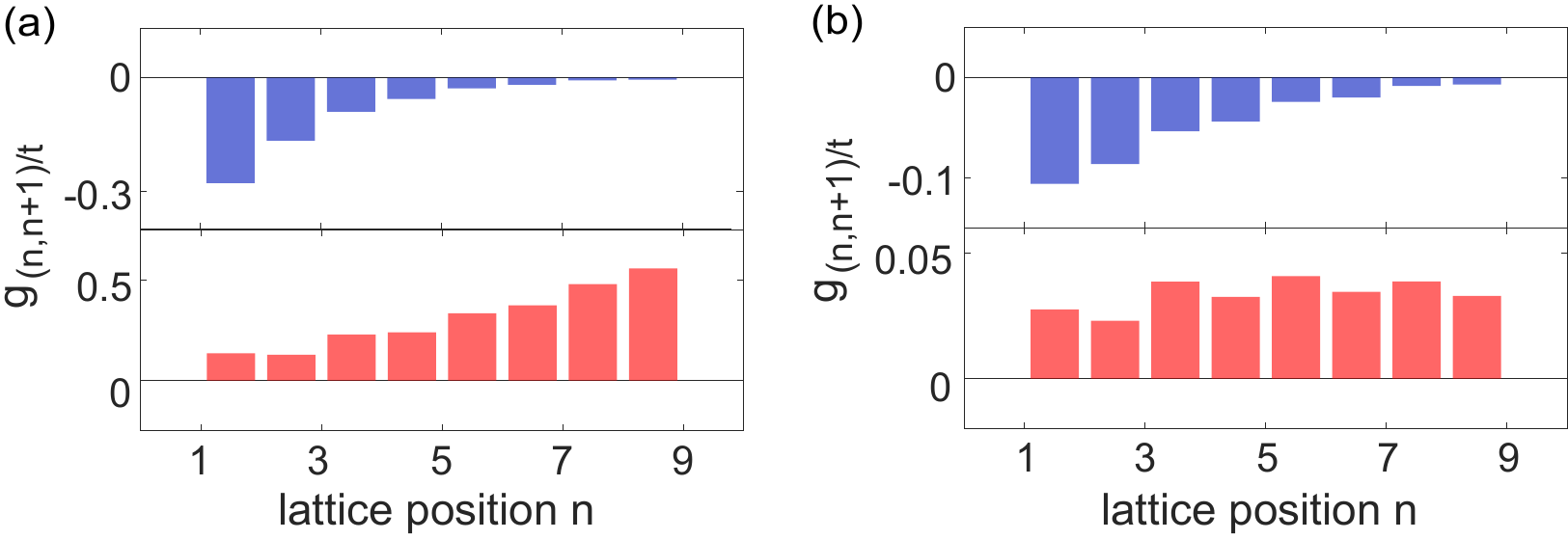}
\caption{\textbf{Power gained or dissipated at coupling junctions in a non-Hermitian gauged laser array at threshold.} Parameters are the same as in Fig.~\ref{fig:gauged} but with $\kappa_0=0.4t$ in (a) and $0.08t$ in (b). The upper (lower) panel shows the result when pumping the leftmost (rightmost) cavity in the tight-binding model.  
} \label{fig:changeKappa}
\end{figure}

As we have mentioned in the main text using the transfer matrix model, selective pumping with an strong cavity loss brings the system to the regime of of zero-mode lasing with tunable spatial profiles, where the lasing mode at threshold has a significant (if not dominating) peak at the pump location. This is also the case in the tight-binding model, but the later does not capture the different mechanisms of 
the anomalous threshold behavior when the physical realization of the imaginary gauge field changes. 

Take the three cases shown in Fig.~\ref{fig:Tmat} analyzed using the transfer matrix approach, for example. They have almost identical asymmetric couplings $t,t'$, and their effective cavity decays, judged by $\im{\omega_\mu}$'s of the passive Hamiltonian $H$, are roughly $1:2:0.4$. The first case gives similar results to the tight-binding model with the same $t'/t$ in Fig.~\ref{fig:gauged} of the main text, where $\kappa_0$ is taken to be $0.2t$. Here we consider the other two cases in the tight-binding model, with $\kappa_0=0.4t$ and $0.08t$ respectively. As can be seen in Fig.~\ref{fig:changeKappa}, the tight-binding model still shows power dissipation (gain) at all coupling junctions when we pump the head (tail) of the non-Hermitian gauged array, and hence it fails to capture the actual power relations displayed in the last two cases in Fig.~\ref{fig:Tmat}.

\begin{figure}[t]
\includegraphics[clip,width=\linewidth]{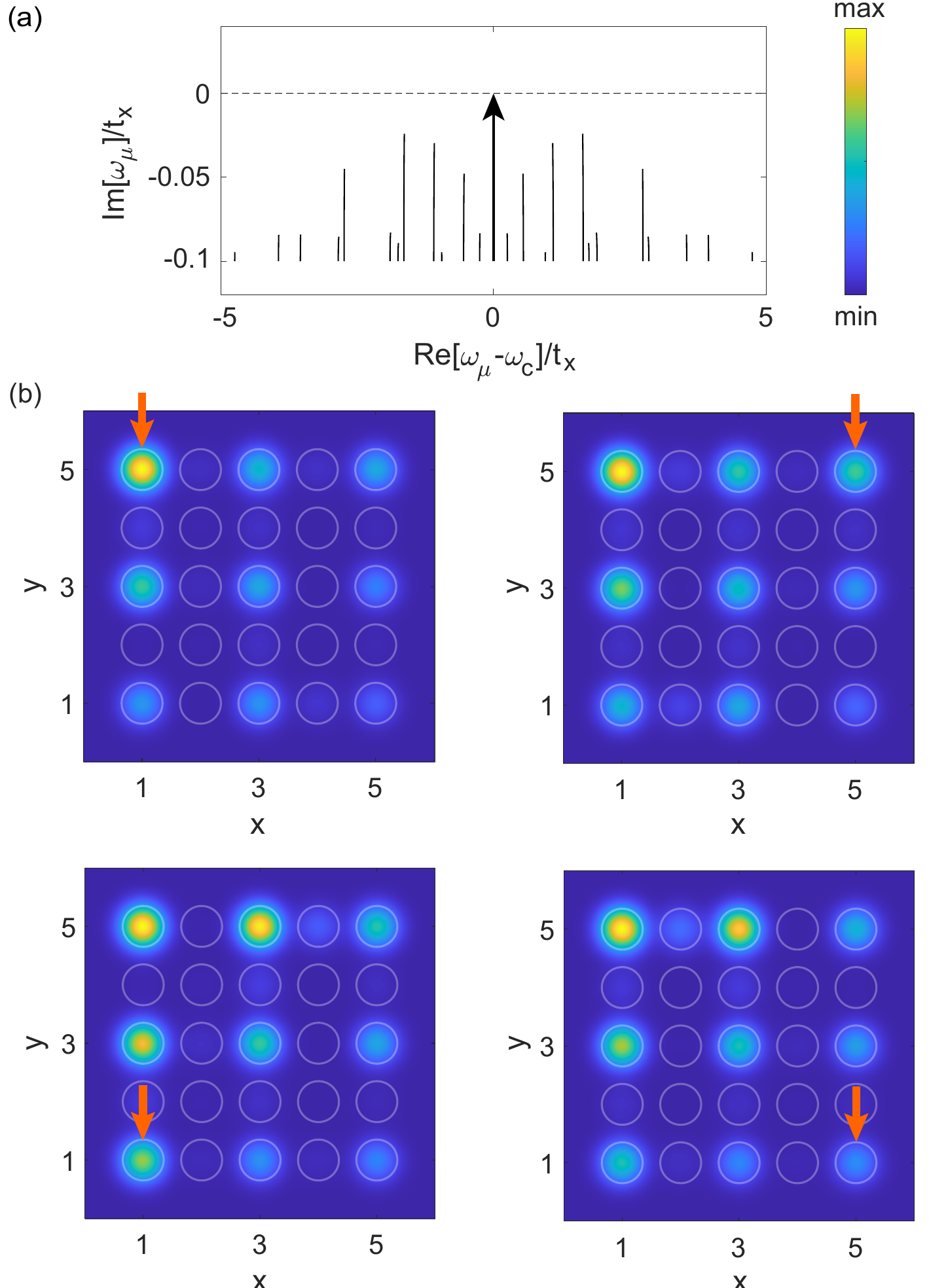}
\caption{\textbf{Selectively pumping a 2D non-Hermitian gauged laser array}. (a) Trajectories of the resonances when pumping either corner of the array. Line with an arrow shows the lasing zero-mode. (b) Field profiles $|\Psi_s|$ of the lasing modes at threshold when pumping either corner (orange arrows), respectively. A Gaussian peak is superposed in each cavity. 
} \label{fig:2dGauged}
\end{figure}

\section{Anomalous threshold behavior in 2D}
\label{sec:2D}

As we have mentioned in the main text, the conventional principle of selective pumping represented by Eq.~(\ref{eq:TH}) in the main text breaks down also in a 2D non-Hermitian gauged array. Here we exemplify in Fig.~\ref{fig:2dGauged} the identical threshold of pumping either corner of a rectangular lattice, which has couplings $t_y=1.5t_x$ and $t'_x/t_x=t'_y/t_y=1.2$. With $\kappa_0=-0.1t_x$, the lasing threshold is reached at $D_sN=8.1\kappa_0$ no matter which corner we pump. The approximation given by Eq.~(\ref{eq:TH_SM}) again captures this behavior qualitatively, giving $D_sN=9\kappa_0$. 

Here we have made the couplings in the $x$ and $y$ directions sufficiently different. If we have not, then the density of states would be high close to the zero-mode, which tend to induce stronger couplings between these non-Hermitian skin modes. As a result, the lasing modes at threshold by pumping the four corners will have significantly difference intensity profiles, even though they still have the same threshold. For example, with the same gauge fields in the $x$ and $y$ directions as above but with $t_y=1.2t_x$, pumping either of the two ``wings'' of the passive zero-mode (i.e., the bottom left and the upper right corners) will introduce strongest intensity peaks at both of these two corners (not shown). Again, this is the regime of zero-mode lasing with tunable spatial profiles mentioned in the main text \cite{zeromodeLaser}. 

\section{Beyond the tight-binding model}
\label{sec:Tmat}

\noindent \textbf{Single ring.} We start by considering a single resonator of refractive index $n$ and length $L$. Its resonances are determined by $e^{inkL} = 1$, or $k=2m\pi/(nL)\,(m=1,2,\ldots)$. Note that $n$ can be made complex with a positive imaginary part to represent both material and radiation losses. Denote $n=n_0+in_1$ and $k=k_r+ik_i$, we find
\be
k_r = \frac{ n_0}{|n|^2} \frac{2m\pi}{L}, \,\,k_i = -\frac{n_1}{|n|^2} \frac{2m\pi}{L}.
\ee

\begin{figure}[b]
\includegraphics[clip,width=\linewidth]{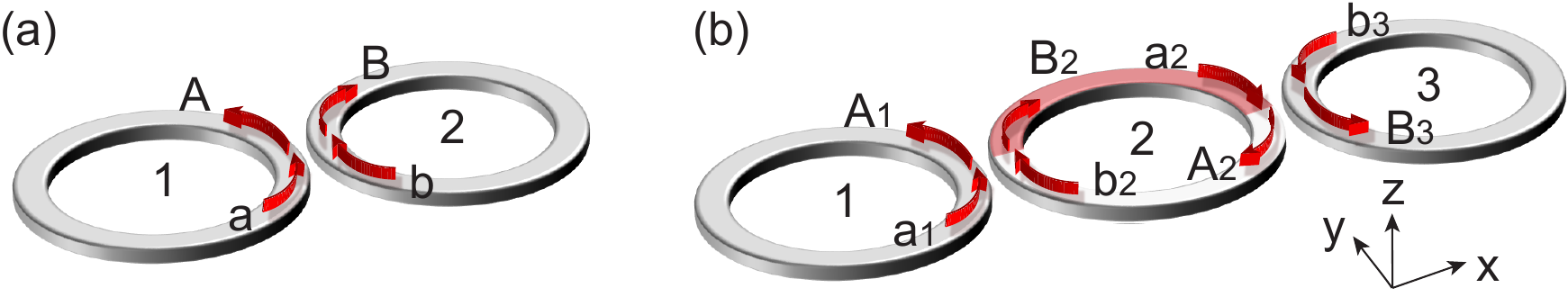}
\caption{Schematics of (a) two and (b) three coupled ring resonators. 
} \label{fig:schematics}
\end{figure}

\noindent \textbf{Two ring with symmetric coupling.} To characterize coupled ring resonators, here we resort to a simple model that ignores the radiative coupling and only considers the evanescent coupling through the scattering matrix $S$ at each coupling junction. We place two identical cavities (1 and 2) next to each other and first consider the coupling between the counterclockwise (CCW) mode in cavity 1 and the clockwise (CW) mode in cavity 2 [Fig.~\ref{fig:schematics}(a)]. 

The incoming and outgoing amplitudes in these two modes at the coupling junction can be captured by the following $S$ matrix \cite{Yariv,inverseVernier}:
\be
\begin{pmatrix}
A\\
B
\end{pmatrix} =
S
\begin{pmatrix}
a\\
b
\end{pmatrix},\quad
S =
\begin{pmatrix}
s & iJ \\
iJ^* & s^*
\end{pmatrix}.\label{eq:S}
\ee
The $S$ matrix is dimensionless, and so are $s$ and $J$. Given the local flux conservation relation, i.e., $|a|^2+|b|^2=|A|^2+|B|^2=[a,b]S^\dagger S[a,b]^T$, $S$ must be unitary ($S^\dagger S=1$) and we have $|s|^2 + |J|^2=1$. 

For a resonance $k$, the amplitudes $a,A$ and $b,B$ are related by
\be
a = Ae^{inkL},\,\, b = Be^{inkL}.
\ee
Substitute these relations in Eq.~(\ref{eq:S}) and we find
\be
(1-se^{inkL})A = iJe^{inkL} B,\,\,(1-s^*e^{inkL})B = iJ^*e^{inkL} A,\label{eq:AB_2rings}
\ee
or
\be
(1-se^{inkL})(1-s^*e^{inkL}) = -|J|^2e^{i2nkL}.
\ee
Using $|s|^2+|J|^2=1$, this expression gives 
\be
e^{i2nkL}-2\re{s}e^{inkL}+1 = 0,\label{eq:exp}
\ee
or
\be
e^{inkL} = e^{\pm i\theta} \label{eq:efac}
\ee
where 
\be
\theta=\tan^{-1}\frac{\sqrt{1-\re{s}^2}}{|\re{s}|}>0 
\ee
given the physical constraint that $|s|\leq1$. 

Equation~(\ref{eq:exp}) indicates that each resonance $k_0$ of a single ring splits into two resonances $k_\pm$ given by
\be
k_\pm-k_0 = \pm\frac{\theta}{nL}.\label{eq:k_pm}
\ee
Due to the mirror symmetry of the system about the $y-z$ plane at the center of the gap between the two rings, these two modes must be symmetric and anti-symmetric about the same plane, i.e., $A=\pm B$ and $a=\pm b$. This physical consideration then determines the phases  of both $s$ and $J$: by substituting the exponential factors in Eq.~(\ref{eq:AB_2rings}) by that given by Eq.~(\ref{eq:exp}), we have
\be
\frac{A}{B} = iJ \frac{e^{\pm i\theta}}{1-se^{\pm i\theta}}\label{eq:A/B}
\ee  
Here we have two choices
\be
iJ \frac{e^{i\theta}}{1-se^{i\theta}} = \pm1,\quad iJ \frac{e^{-i\theta}}{1-se^{-i\theta}} = \mp1,
\ee  
that give
\be
(s\pm iJ)e^{i\theta} = 1,\quad (s\mp iJ)e^{-i\theta} = 1,
\ee
respectively. No matter which choice we make,  
\be
s^2+J^2=1
\ee
always holds, and combining it with the local flux conservation relation $|s|^2+|J|^2=1$, we know that both $s$ and $J$ are real. Furthermore, we do not expect a phase jump when $a$ ($b$) passes through the coupling junction to become part of $A$ ($B$), especially in the limit $J\rightarrow0$: we should recover the single-ring case presented above, i.e., $s\rightarrow1$. Therefore, we take $s$ to be positive, and the angle $\theta$ becomes $\theta=\tan^{-1}(|J|/s)$.

The two choices in Eq.~(\ref{eq:A/B}) then tells us that if $J$ is also positive, then $k_-$ is symmetric and $k_+$ is anti-symmetric. If $J$ is negative instead, then $k_+$ is symmetric and $k_-$ is anti-symmetric.

Equation~(\ref{eq:k_pm}) indicates that 
\be
\tilde{t}\equiv k_+-k_0=k_0-k_-=\frac{\theta}{nL}
\ee
is the effective coupling between the CCW mode in cavity 1 and CW mode in cavity 2 in the tight-binding model. It is approximately real for a high-Q passive resonance (i.e., with $0<n_1\ll n_0$). 
$|\tilde{t}|$ approaches its maximum $\pi/(2n L)=k_0/(4m)$, i.e., the the strong coupling limit, when $|J|\gg s$. 

Finally, the analysis for the coupling of the CW mode in cavity 1 and the CCW mode in cavity 2 is the same as above, leading to the same resonances $k_\pm$. Altogether, there are two pairs of degenerate modes near $k_0$, one symmetric and the other anti-symmetric about the $y-z$ mirror plane. Because the system is also mirror symmetric about the $x-z$ plane through the centers of the two rings, the two modes in each pair can also be expressed as symmetric (cosine) and anti-symmetric (sine) modes about this plane.\\ 

\noindent \textbf{Two rings with asymmetric couplings.} In the main text, we have introduced the transfer matrix to analyze a coupled laser array with cavity rings and auxiliary rings. Here we study a special case with just three rings [Fig.~\ref{fig:schematics}(b)]. The second ring is the auxiliary ring, and it features different values of the refractive index and arclengths in the upper ($n_{u}$,$L_{u}$) and lower ($n_d$,$L_{d}$) halves, allowing us to implement different combinations of gain and loss. We also allow the refractive index to be different in the left ($n_1$) and right ($n_3$) rings, to account for selectively pumping one ring. The coupling junctions between the first two and the last two are identical, and so are their scattering matrices:
\be
\begin{pmatrix}
A_1\\
B_2
\end{pmatrix} =
S
\begin{pmatrix}
a_1\\
b_2
\end{pmatrix},\quad
\begin{pmatrix}
B_3\\
A_2
\end{pmatrix} =
S
\begin{pmatrix}
b_3\\
a_2
\end{pmatrix}.
\ee
We take $s>0$ and $J\in\mathbb{R}$ in the $S$ matrix using our findings of the two ring case. We also have $a_1 = A_1e^{in_1kL}\equiv A_1e_1$, $b_2 = A_2e^{in_dkL_d}\equiv A_2e_d$, $a_2 = B_2e^{in_ukL_u}\equiv B_2 e_u$, $b_3 = B_3e^{in_3kL}\equiv B_3e_3$, 
which leads to
\be
\frac{s-e_1}{1-se_1}\frac{s-e_3}{1-se_3}e_u e_d = 1 \label{eq:3rings_sm}
\ee
when we combine the two scattering matrix equations. We note that $J$ does not appear in Eq.~(\ref{eq:3rings_sm}) similar to Eq.~(\ref{eq:exp}) in the two-ring case; it is eliminated in the coefficients of $e_1$ and $e_3$ in the numerator, both of which are given by $s^2+J^2=1$. 

Obviously, Eq.~(\ref{eq:3rings_sm}) is equivalent to Eq.~(\ref{eq:Nrings}) in the main text with $N=3$, as can be checked directly using 
\be
M = 
\frac{1}{J^2}
\begin{bmatrix}
s(e_d - e_u^{-1}) & s^2e_u^{-1} - e_d \\
s^2e_d - e_u^{-1} & s(e_u^{-1}-e_d)
\end{bmatrix}.
\ee
We have mentioned the explicit no-Hermitian gauge symmetry exhibited by Eq.~(\ref{eq:Nrings}) in the main text. As expected, here Eq.~(\ref{eq:3rings_sm}) shows the same non-Hermitian gauge symmetry: $n_u,L_u$ and $n_d,L_d$ only appear in the product $e_ue_d$. Therefore, as long as 
\be
\sigma \equiv n_uL_u+n_dL_d = const.,
\ee 
the coupled rings have the same resonances for a fixed set of other parameters, independent of the values of $n_u$,$n_d$, and in particular, their imaginary parts. 


There are several scenarios possible due to this non-Hermitian gauge symmetry: (1) With net loss ($\im{\sigma}>0$), an imaginary gauge transformation can map an auxiliary ring with different losses in the two halves to one with the same loss. It can also map an auxiliary ring with gain and loss halves to one with stronger gain and stronger loss, or to one with weaker loss(es) in the two halves. (2) With net gain ($\im{\sigma}<0$), an imaginary gauge transformation can map an auxiliary ring with different gains in the two halves to one with the same gain. It can also map an auxiliary ring with gain and loss halves to one with stronger gain and stronger loss, or to one with weaker gain(s) in the two halves.

Below we analyze Eq.~(\ref{eq:3rings_sm}) to estimate the effective couplings and the non-Hermitian gauge field. When $s=1$ (and $J=0$), the rings are uncoupled and Eq.~(\ref{eq:3rings_sm}) only gives the condition that determines the resonances of the auxiliary ring, i.e., $e_ue_d=1$; the resonances of cavities 1 and 2 are determined by $e_1=1$ and $e_2=1$ respectively, which make the denominators (and numerators) of the two fractions on the left hand side of Eq.~(\ref{eq:3rings_sm}) vanish. From this analysis, we expect three ``supermodes'' from Eq.~(\ref{eq:3rings_sm}) when $|J|\neq0$, which evolve from three uncoupled resonances in the three rings, one from each. 

\begin{figure}[b]
\includegraphics[clip,width=\linewidth]{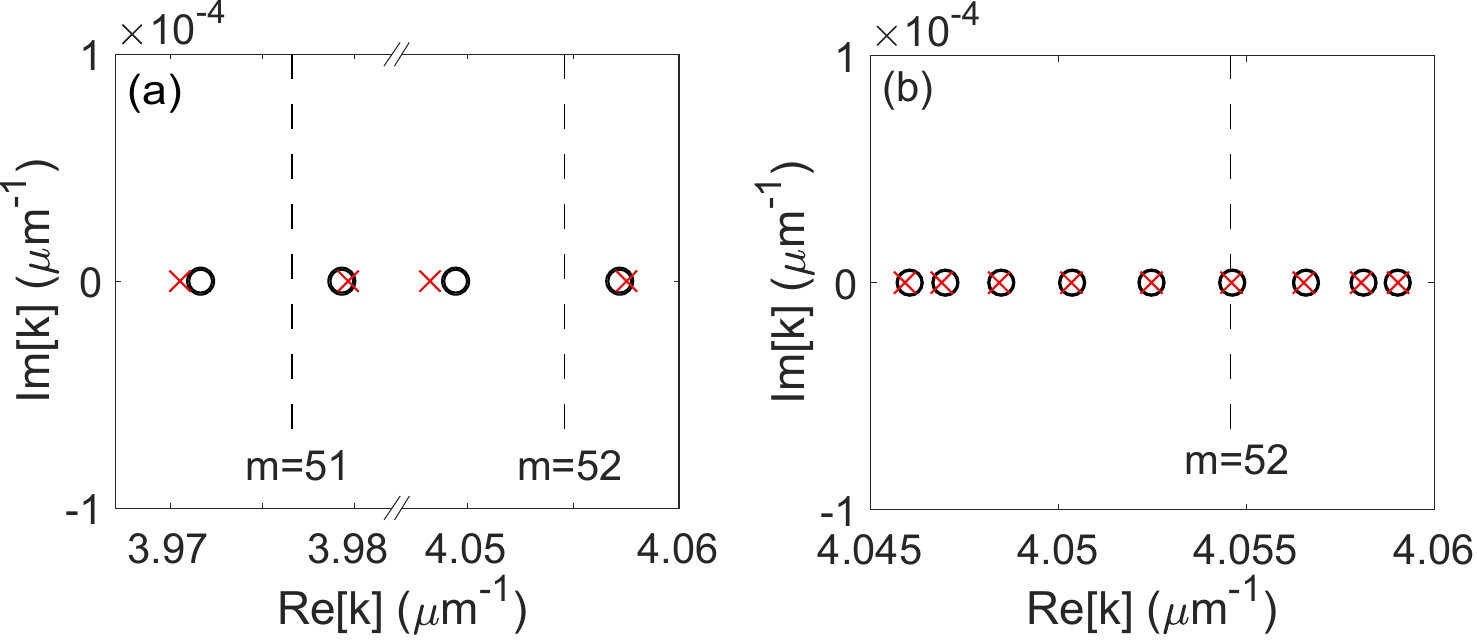}
\caption{Real-valued resonances in (a) a three-ring system and (b) a 17-ring system. Parameter used are $s=0.7$, $L=29.845\,\mu{m}\,(r=4.75\,\mu{m})$, $L_u=L_d=15.101\,\mu{m}\,(r=4.81\,\mu{m})$, and all rings have $n=2.7$ with loss ignored. Legend: single-ring resonances (dashed lines), numerical results of coupled resonances (circles), and their approximations (crosses).   
} \label{fig:comparing_couplings}
\end{figure}

When cavities 1 and 2 are identical, the effective coupling between them can be expressed by reformatting Eq.~(\ref{eq:3rings_sm}):
\be
k-k_0 = \frac{1}{in_1L}\ln\frac{1\pm s\sqrt{e_u(k)e_d(k)}}{s\pm \sqrt{e_u(k)e_d(k)}}. \label{eq:3rings_couplingConstant}
\ee
We have included the explicit $k$ dependence of $e_u$,$e_d$ to emphasize that this is in fact a self-consistent equation. The two signs on the right hand need to be taken as the same, which leads to two resonances $k_\pm$ as a result of the effective coupling between cavities 1 and 2. When this effective coupling is much weaker than the free spectral range of the single-ring resonances, $k_\pm$ can be approximated by 
\be
k_\pm = k_0 + \frac{1}{in_1L}\ln\frac{1\pm s\sqrt{e_u(k_0)e_d(k_0)}}{s\pm \sqrt{e_u(k_0)e_d(k_0)}}. \label{eq:3rings_couplingConstant2}
\ee
We note that unlike the two-ring case, $|k_\pm-k_0|$ here are different in general, even when the system is idealized to be Hermitian [see Fig.~\ref{fig:comparing_couplings}(a)], which seems to suggest that an effective symmetric coupling or asymmetric couplings would be difficult to define. As shown in Fig.~\ref{fig:comparing_couplings}(b) for a longer array with 17 rings (nine cavity rings plus eight auxiliary rings) though, this issue does not impose a fundamental challenge when comparing the results of the tight-binding model and the transfer matrix approach: if we define the effective coupling in the three-ring case as $\tilde{t}=(k_+ - k_-)/2$ using Eq.~(\ref{eq:3rings_couplingConstant}), and that in the 17-ring case by fitting the resonances with a tight-binding model ($\tilde{t}=\sqrt{tt'}$), they differ by about $11\%$, giving by $\tilde{t}=3.88\times 10^{-3}\,\mu{m}^{-1}$ and $3.45\times 10^{-3} \,\mu{m}^{-1}$. 

Furthermore, by tuning the size of the auxiliary rings to be anti-resonant with the cavity ring, we can align the center of the supermodes with the single-ring resonance $k$. This can be seen by requiring that the two solutions of Eq.~(\ref{eq:3rings_couplingConstant}), i.e., $k_\pm$, satisfy $k_+-k_0 = - (k_--k_0) = \tilde{t}$, or equivalently,
\be
\frac{1+ s\sqrt{e_u(k_+)e_d(k_+)}}{s+ \sqrt{e_u(k_+)e_d(k_+)}}\frac{1- s\sqrt{e_u(k_-)e_d(k_-)}}{s- \sqrt{e_u(k_-)e_d(k_-)}}=1.\label{eq:antiResonant}
\ee
We then find 
\be
\sqrt{e_u(k_+)e_d(k_+)e_u(k_-)e_d(k_-)} = -1.
\ee
Next, we substitute $k_\pm = k_0 \pm \tilde{t}$ into this expression and use $e_u(k_+)e_u(k_-) = e_u^2(k_0)$, $e_d(k_+)e_d(k_-) = e_d^2(k_0)$, we then find
\be
e_u(k_0)e_d(k_0) = \pm 1.
\ee

\begin{figure}[t]
\includegraphics[clip,width=0.95\linewidth]{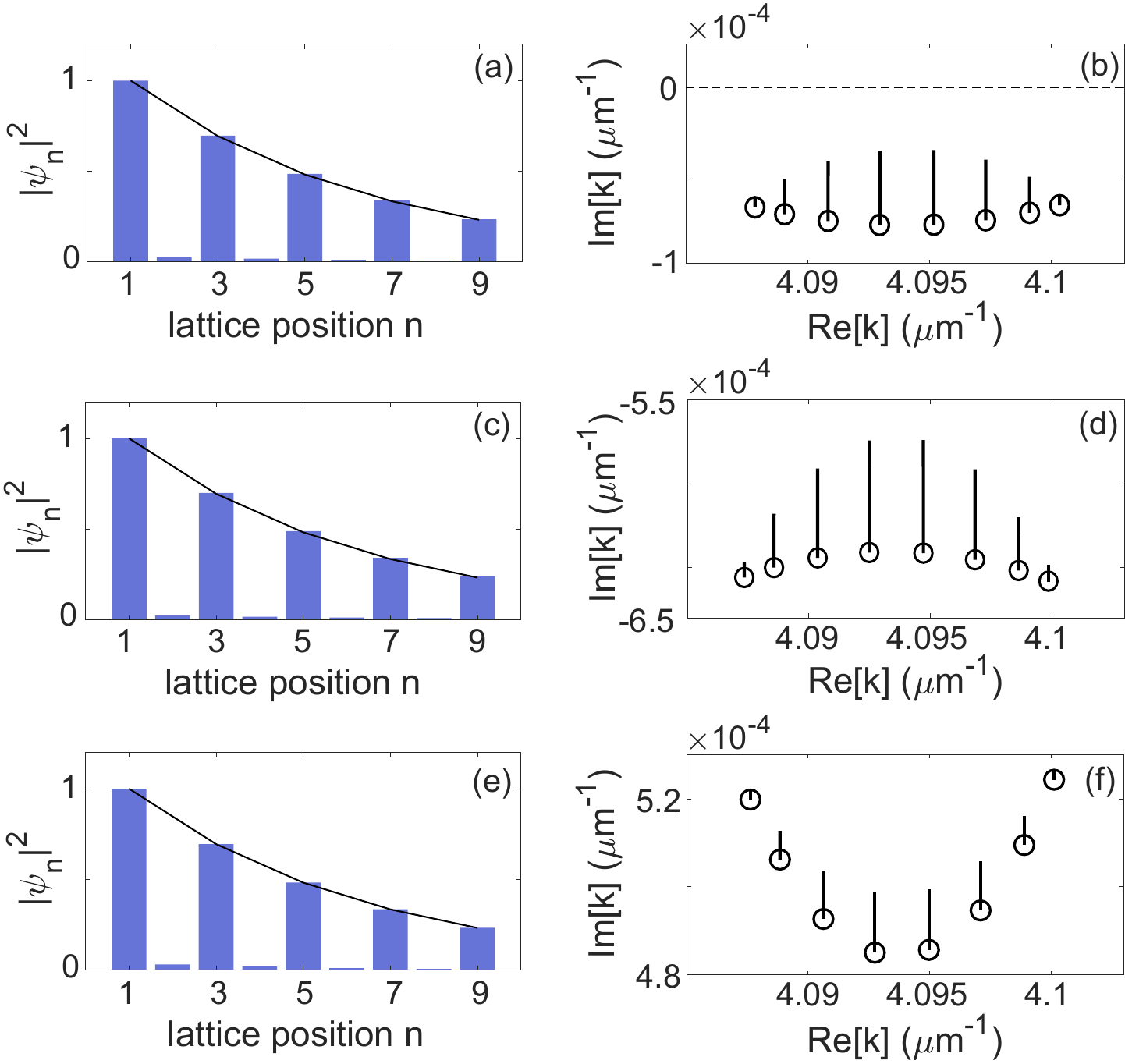}
\caption{\textbf{Supplemental data for the three cases shown in Fig.~\ref{fig:Tmat} of the main text.}
(a,c,e) Spatial profiles of their passive zero-mode from the transfer matrix calculation (histogram), plotted using $\Psi_n = A_{2n-1}\,(n=1,2,\ldots,9)$ of the CCW modes in the cavity rings. Solid line shows their envelope from tight-binding model with $t'/t=1.2$. (b,d,f) Trajectories of auxiliary-ring resonances when the pump is increased to the lasing threshold of the non-Hermitian skin zero-mode. Open dots mark the passive resonances.
} \label{fig:17ring_SM}
\end{figure}

\noindent The ``$+$'' solution is spurious as can be checked when substituted in Eq.~(\ref{eq:antiResonant}), and the ``$-$'' solution is precisely the anti-resonant condition in an auxiliary ring. Here this solution requires $L_{u,d}=14.779\,\mu{m}$, which gives $\tilde{t}=3.74\times 10^{-3}\,\mu{m}^{-1}$ in the three-ring case. 

We note that the estimation of the gauge field is independent from that of $\tilde{t}$: the latter, as mentioned, is given by $\sqrt{tt'}$, and the former is given by $|t'/t|$. In the tight-binding model for two asymmetrically coupled rings, the amplitude ratio between the rings are $\pm\sqrt{t'/t}$. In the transfer matrix analysis, this ratio is represented by ${A_1}/{B_3}$ that can be shown to be  
\be
\frac{A_1}{B_3} = e_d\frac{s-e_3}{1-se_1} = e_u^{-1}\frac{1-se_3}{s-e}.
\ee
In the last step we have used Eq.~(\ref{eq:3rings_sm}). Using $e_1=e_3$ in the passive case (and with uniform pumping), we then find
\be
\left|\frac{t'}{t}\right| = \left|\frac{A_1}{B_3}\right|^2 = \left|\frac{e_d(k_\pm)}{e_u(k_\pm)}\right|\approx\left|\frac{e_d(k_0)}{e_u(k_0)}\right|\approx e^{\im{n_uL_u-n_dL_d}k_0}.\label{eq:gaugeField}
\ee
It gives $|t'/t|\approx1.20$ for the three sets of parameters used in Fig.~\ref{fig:Tmat}, which agree nicely with the values of the gauge field extracted from the 17 rings case (i.e., $|t'/t|\approx1.20$). 

The analysis for the couplings of CW modes in cavities 1 and 2 is slightly different; the one in cavity 1 (cavity 2) couples to the bottom (upper) half of the auxiliary ring, and hence $e_u$ and $e_d$ need to be switched when compared with the analysis above for the CCW modes in these two cavities. Nevertheless, this change does not affect the resonances due to the non-Hermitian gauge symmetry mentioned earlier. Therefore, here we have three pairs of degenerate supermodes with all modes considered (two pairs originated from the cavity rings and one pair from the auxiliary ring). This observation holds for both uniform pumping and selective pumping (i.e., different $n_i$'s in the first and last rings).
\\

\noindent \textbf{Longer arrays.} In the main text, we have used the expression (\ref{eq:gaugeField}) to keep the non-Hermitian gauge field fixed when varying the gain and loss in the two halves of auxiliary rings. In Figs.~\ref{fig:17ring_SM}(a,c,e) we show that this approximation has a high precision by comparing the amplitudes of the CCW zero-mode in the cavity rings with their envelope from the tight-binding model. In Figs.~\ref{fig:17ring_SM}(b,d,f) we show the spectra of the supermodes formed by the auxiliary ring resonances, which do not lase before the non-Hermitian skin zero-mode except for the last case.


\begin{thebibliography}{99}


\bibitem{Hermite} H. Laabs and B. Ozygus, \textit{Excitation of Hermite Gaussian modes in end-pumped solid-state lasers via off-axis pumping}, Opt. Laser Technol. \textbf{28}, 213--214 (1996).
\bibitem{Hermite2} Y. F. Chen, T. M. Huang, C. F. Kao, C. L. Wang, and S. C. Wang, \textit{Generation of Hermite-Gaussian modes in fiber-coupled laser-diode end-pumped lasers}, IEEE J. Quan. Electron. \textbf{33}, 1025--1031 (1997).
\bibitem{Woerdman} J. Dingjan, M. P. van Exter, and J. P. Woerdman, \textit{Geometric modes in a single-frequency Nd: YVO 4 laser},
Opt. comm. \textbf{188}, 345--351 (2001).


\bibitem{Narimanov} S. Shinohara, T. Harayama, T. Fukushima, M. Hentschel, T. Sasaki, and E. E. Narimanov, 
\textit{Chaos-assisted directional light emission from microcavity lasers}, Phys. Rev. Lett. \textbf{104}, 163902 (2010).

\bibitem{Wang} S. Wang, S. Liu, Y. Liu, S. Xiao, Z. Wang, Y. Fan, J. Han, L. Ge, and Q. Song, \textit{Direct Observation of Chaotic Resonances in Optical Microcavities}, Light Sci. Appl. \textbf{10}, 135 (2021).


\bibitem{rex}N. B. Rex, R. K. Chang, and L. J. Guido, \textit{Threshold lowering in GaN micropillar lasers by means of spatially selective optical pumping}, IEEE Photonics Technology Lett. {\bf 13} 1 (2002).
\bibitem{Aung} N. L. Aung, L. Ge, O. Malik, H. E. T\"ureci, and C. F. Gmachl, \textit{Threshold Current Reduction and Directional Emission of Deformed Microdisk Lasers via Spatially Selective Electrical Pumping}, Appl. Phys. Lett. \textbf{107}, 151106 (2015).
\bibitem{Ge_selective1} L. Ge, O. Malik, and H. E. T\"ureci, \textit{Enhancement of laser powerefficiency by control of spatial hole burning interactions}, Nat. Photon. \textbf{8}, 871 (2014).

\bibitem{spiral} G. D. Chern, H. E. T\"ureci, A. Douglas Stone, R. K. Chang, M. Kneissl, and N. M. Johnson, \textit{Unidirectional lasing from InGaN multiple-quantum-well spiral-shaped micropillars}, Appl. Phys. Lett. {\bf 83}, 1710 (2003).
\bibitem{Choi} M. Choi, T. Tanaka, T. Fukushima, and T. Harayama, \textit{Control of directional emission in quasistadium microcavity laser diodes with two electrodes}, Appl. Phys. Lett. 88, 211110 (2006).
\bibitem{Seng} S. F. Liew, B. Redding, L. Ge, G. S. Solomon, and H. Cao, \textit{Active control of emission directionality of semiconductor microdisk lasers}, App. Phys. Lett. \textbf{104}, 231108 (2014).
\bibitem{Fukushima} T. Fukushima, T. Harayama, P. Davis, P. O. Vaccaro, T. Nishimura, and T. Aida, \textit{Ring and axis mode lasing in quasi-stadium laser diodes with concentric end mirrors}, Opt. Lett. \textbf{27}, 1430-1432 (2002).

\bibitem{Ge_selective2} L. Ge, \textit{Selective Excitation of Lasing Modes by Controlling Modal Interactions}, Opt. Express \textbf{23}, 30049 (2015).
\bibitem{Seng2} S. F. Liew, L. Ge, B. Redding, S. Solomon, and H. Cao, \textit{Pump-controlled modal interactions in microdisk lasers}, Phys. Rev. A \textbf{91}, 043828 (2015).
\bibitem{Ge_selective3} L. Ge, H. Cao, and A. D. Stone, \textit{Condensation of Thresholds in Multimode Microlasers}, 
Phys. Rev. A \textbf{95}, 023842 (2017).

\bibitem{Rotter} T. Hisch, M. Liertzer, D. Pogany, F. Mintert, and S. Rotter, \textit{Pump-controlled directional light emission from random lasers}, Phys. Rev. Lett. {\bf 111}, 023902 (2013).
\bibitem{Sebbah} N. Bachelard, J. Andreasen, S. Gigan, and P. Sebbah, \textit{Taming random lasers through active spatial control of the pump}, Phys. Rev. Lett. {\bf 109}, 033903 (2012).
\bibitem{Sebbah2} N. Bachelard, S. Gigan, X. Noblin, and P. Sebbah, \textit{Adaptive pumping for spectral control of random lasers},
Nat. Phys. \textbf{10}, 426 (2014).

\bibitem{Manni} F. Manni, K. G. Lagoudakis, T. C. H. Liew, R. Andr\'e, and B. Deveaud-Pl\'edran, \textit{Spontaneous Pattern Formation in a Polariton Condensate}, Phys. Rev. Lett. {\bf 107}, 106401 (2011).
\bibitem{Ge_selective4} L. Ge, A. Nersisyan, B. Oztop, and H. E. T\"ureci, \textit{Pattern Formation and Strong Nonlinear Interactions in Exciton-Polariton Condensates}, arXiv:1311.4847 (2013).
\bibitem{Sun} Y. Sun, Y. Yoon, S. Khan, L. Ge, M. Steger, L. N. Pfeiffer, K. West, H. E. T\"ureci, D. W. Snoke, and K. A. Nelson, \textit{Stable Switching among High-Order Modes in Polariton Condensates}, Phys. Rev. B \textbf{97}, 045303 (2018).

\bibitem{Gao} Z. Gao, M. T. Johnson, and K. D. Choquette, \textit{Rate Equation Analysis and Non-Hermiticity in Coupled Semiconductor Laser Arrays}, J. Appl. Phys. \textbf{123}, 173102 (2018).

\bibitem{Dave} H. Dave, Z. Gao, S. T. M. Fryslie, B. J. Thompson, and K. D. Choquette, \textit{Static and Dynamic Properties of Coherently-Coupled Photonic-Crystal Vertical-Cavity Surface-Emitting Laser Arrays}, IEEE J. Sel. Top. Quantum Electron. \textbf{25}, 1 (2019).

\bibitem{Kominis} Y. Kominis, V. Kovanis, and T. Bountis, \textit{Controllable Asymmetric Phase-Locked States of the Fundamental Active Photonic Dimer}, Phys. Rev. A \textbf{96}, 043836 (2017).

\bibitem{Kominis2} Y. Kominis, A. Bountis, and V. Kovanis, \textit{Radically Tunable Ultrafast Photonic Oscillators via Differential Pumping}, J. Appl. Phys. \text{127}, 083103 (2020).

\bibitem{Longhi} S. Longhi, Y. Kominis, and V. Kovanis, \textit{Presence of Temporal Dynamical Instabilities in Topological Insulator Lasers}, Europhys. Lett. \textbf{122}, 1 (2018).


\bibitem{Microcavity1} \textit{Optical Processes in Microcavities}, edited by R. K. Chang and A. J. Campillo, Advanced Series in Applied Physics (World Scientific, Singapore, 1996).
\bibitem{Microcavity2} \textit{Optical Microcavities}, edited by K. J. Vahala, Advanced Series in Applied Physics (World Scientific, Singapore, 2004).
\bibitem{Microcavity3} H. Cao and J. Wiersig, \textit{Dielectric microcavities: Model systems for wave chaos and non-Hermitian physics},
Rev. Mod. Phys. \textbf{87}, 61--111 (2015).

\bibitem{RMP} V. V. Konotop, J. Yang, and D. A. Zezyulin, \textit{Nonlinear waves in PT-symmetric systems,} Rev. Mod. Phys. \textbf{88}, 035002 (2016).
\bibitem{NPreview} L. Feng, R. El-Ganainy, and L. Ge, \textit{Non-Hermitian photonics based on parity-time symmetry}, Nat. Photonics \textbf{11}, 752--762 (2017).
\bibitem{NPhyreview} R. El-Ganainy, K. G. Makris, M. Khajavikhan, Z. H. Musslimani, S. Rotter, and D. N. Christodoulides, \textit{Non-Hermitian physics and PT symmetry}, Nat. Phys. \textbf{14}, 11--19 (2018).
\bibitem{NMatreview} \c{S}. K. \"{O}zdemir, S. Rotter, F. Nori, and L. Yang, \textit{Parity–Time Symmetry and Exceptional Points in Photonics}, Nat. Mater. \textbf{18}, 783 (2019).



\bibitem{Florent} F. Baboux, L. Ge, T. Jacqmin, M. Biondi, E. Galopin, A. Lema\^{i}tre, L. Le Gratiet, I. Sagnes, S. Schmidt, H. E. T\"ureci, A. Amo, and J. Bloch, \textit{Bosonic Condensation and Disorder-Induced Localization in a Flat Band}, Phys. Rev. Lett. \textbf{116}, 066402 (2016).
\bibitem{Ge_selective5} L. Ge, \textit{Parity-Time Symmetry in a Flat-Band System}, Phys. Rev. A \textbf{92}, 052103 (2015).

\bibitem{NHFlatband_PRL} B. Qi, L. Zhang and L. Ge, \textit{Defect states emerging from a non-Hermitian flat band of photonic zero modes},
Phy. Rev. Lett. \textbf{120}, 093901 (2018).
\bibitem{NHFlatband_PRJ} L. Ge, \textit{Non-Hermitian lattices with a flat band and polynomial power increase},
Photon. Res. \textbf{6}, A10--A17 (2018).

\bibitem{zeromodeLaser} L. Ge, \textit{Symmetry-protected zero-mode laser with a tunable spatial profile},
Phys. Rev. A \textbf{95}, 023812 (2017).

\bibitem{Zhao} H. Zhao, P. Miao, M. H. Teimourpour, S. Malzard, R. El-Ganainy, H. Schomerus, and L. Feng, \textit{Topological hybrid silicon microlasers}, Nat. Commun. \textbf{9}, 981 (2018).
\bibitem{Poli} C. Poli, M. Bellec, U. Kuhl, F. Mortessagne, and H. Schomerus, \textit{Selective enhancement of topologically induced interface states in a dielectric resonator chain}, Nat. Commun. \textbf{6}, 6710 (2015).
\bibitem{St-Jean} P. St-Jean, V. Goblot, E. Galopin, A. Lemaître, T. Ozawa, L. Le Gratiet, I. Sagnes, J. Bloch, and A. Amo, \textit{Lasing in topological edge states of a one-dimensional lattice}, Nat. Photon. \textbf{11}, 651--656 (2017).

\bibitem{Bandres} M. A. Bandres, S. Wittek, G. Harari, M. Parto, J. Ren, and M. Segev, \textit{Topological insulator laser: Experiments}, 
Science \textbf{359}, eaar4005 (2018).


\bibitem{steering} H. Zhao, X. Qiao, T. Wu, B. Midya, S. Longhi, and L. Feng, \textit{Non-Hermitian Topological Light Steering}, 
Science \textbf{365}, 1163 (2019).


\bibitem{Hatano} N. Hatano and D. R. Nelson, \textit{Localization Transitions in Non-Hermitian Quantum Mechanics,}
Phys. Rev. Lett. \textbf{77}, 570--573 (1996).
\bibitem{Longhi_gauge} S. Longhi, D. Gatti, and G. D. Valle, \textit{Robust Light Transport in Non-Hermitian Photonic Lattices}, 
Sci. Reports \textbf{5}, 13376 (2015).
\bibitem{Song_gauge} F. Song, S. Yao, and Z. Wang, \textit{Non-Hermitian Skin Effect and Chiral Damping in Open Quantum Systems}, 
Phys. Rev. Lett. \textbf{123}, 170401 (2019).
\bibitem{Szameit_gauge} S. Weidemann, M. Kremer, T. Helbig, T. Hofmann, A. Stegmaier, M. Greiter, R. Thomale, and A. Szameit, \textit{Topological Funneling of Light}, Science \textbf{368}, 311 (2020).
\bibitem{Li_gauge} L. Li, C. H. Lee, and J. Gong, \textit{Topological Switch for Non-Hermitian Skin Effect in Cold-Atom Systems with Loss}, Phys. Rev. Lett. \textbf{124}, 250402 (2020).
\bibitem{Zhang_gauge} L. Zhang \textit{et al.}, \textit{Acoustic non-Hermitian skin effect from twisted winding topology}, 
Nat. Commun. \textbf{12}, 6297 (2021).
\bibitem{Wang_Nature} W. Wang, X. Wang, and G. Ma, \textit{Non-Hermitian Morphing of Topological Modes}, 
Nature \textbf{608}, 7921 (2022).
\bibitem{Ge_gauge} J. D. H. Rivero and L. Ge, \textit{Chiral Symmetry in Non-Hermitian Systems: Product Rule and Clifford Algebra}, 
Phys. Rev. B \textbf{103}, 014111 (2021).
\bibitem{Feng_gauge} Z. Zhang \textit{et al.}, \textit{Tunable Topological Charge Vortex Microlaser}, 
Science \textbf{368}, 760 (2020).
\bibitem{Feng_gauge2} Z. Zhang et al., \textit{Spin–Orbit Microlaser Emitting in a Four-Dimensional Hilbert Space}, 
Nature \textbf{612}, 246 (2022).
\bibitem{Feng_gauge3} Z. Gao, X. Qiao, M. Pan, S. Wu, J. Yim, K. Chen, B. Midya, L. Ge, and L. Feng, \textit{Two-dimensional reconfigurable non-Hermitian gauged laser array}, Phys. Rev. Lett. \textbf{130}, 263801 (2023).

\bibitem{SSH} W. P. Su, J. R. Schrieffer, and A. J. Heeger, \textit{Solitons in polyacetylene},
Phys. Rev. Lett. \textbf{42}, 1698 (1979).

\bibitem{bibnote:2} Note $p_1 = {\cal J}_{2,1}$ and $p_{N} = {\cal J}_{N,N-1}$ at the two ends. 

\bibitem{bibnote:1} Here we have assumed that the lasing threshold is not at an exceptional point, which would make $\tilde{\Psi}^T_0\Psi_0=0$.

\bibitem{Rivero_PRL_2022} J. H. D. Rivero, L. Feng, and L. Ge, \textit{Imaginary Gauge Transformation in Momentum Space and Dirac Exceptional Point}, Phys. Rev. Lett. \textbf{129}, 243901 (2022).

\bibitem{Ge_PRA_2017b} L. Ge, K. G. Makris, and L. Zhang, \textit{Optical Fluxes in Coupled PT-Symmetric Photonic Structures}, 
Phys. Rev. A \textbf{96}, 023820 (2017).

 

\bibitem{Yariv} A.~Yariv, \textit{Critical coupling and its control in optical waveguide-ring resonator systems}, Electron. Lett. {\bf 36}, 321-322 (2000).
\bibitem{inverseVernier} L. Ge and H. E. T\"ureci, \textit{Inverse Vernier Effect in Coupled Lasers}, 
Phys. Rev. A \textbf{92}, 013840 (2015).

\bibitem{SALT_Science} H. E. T\"ureci, L. Ge, S. Rotter, and A. D. Stone, \textit{Storng interactions in multimode random lasers},
Science {\bf 320}, 643 (2008).

\bibitem{SALT_Nonlinearity} H. E. T\"ureci, A. D. Stone, L. Ge, S. Rotter, and R. J. Tandy, \textit{Ab Initio Self-Consistent Laser Theory and Random Lasers}, Nonlinearity \textbf{22}, C1 (2008).


\bibitem{Ge_SciRep} L. Ge and R. El-Ganainy, \textit{Nonlinear Modal Interactions in Parity-Time (PT) Symmetric Lasers}, 
Sci. Rep. \textbf{6}, 24889 (2016).

\end{thebibliography}
\end{document}